\documentclass[journal]{IEEEtran}
\UseRawInputEncoding
\usepackage{url}
\usepackage{amsmath,amsfonts,amssymb}
\usepackage{tabularx}
\usepackage{graphicx}
\usepackage{comment}
\usepackage{color}
\usepackage[acronym]{glossaries}
\usepackage{hyperref}
\usepackage{lineno}
\usepackage{subcaption}
\usepackage{algorithm}
\usepackage{algorithmic}
\usepackage{stfloats}
\usepackage{caption}
\usepackage{soul}
\usepackage{xspace}
\usepackage{caption}
\newacronym{5g}{5G}{Fifth Generation of mobile networks}
\newacronym{4g}{4G}{Fourth generation of mobile networks}
\newacronym{6g}{6G}{Sixth generation of mobile networks}
\newacronym{ai}{AI}{Artificial Intelligence}
\newacronym[shortplural={RUs},longplural={Radio Units}]{ru}{RU}{Radio Unit}
\newacronym[shortplural={CUs},longplural={Centralized Units}]{cu}{CU}{Centralized Unit}
\newacronym[shortplural={O-RUs},longplural={Open Radio Units}]{oru}{O-RU}{Radio Unit}
\newacronym[shortplural={O-CUs},longplural={Open Centralized Units}]{ocu}{O-CU}{Open Centralized Unit}
\newacronym[shortplural={CU-CPs},longplural={Centralized Units Control Plane}]{cucp}{CU-CP}{Central Unit Control Plane}
\newacronym[shortplural={CU-UPs},longplural={Centralized Units User Plane}]{cuup}{CU-UP}{Central Unit User Plane}
\newacronym[shortplural={DUs},longplural={Distributed Units}]{du}{DU}{Distributed Unit}
\newacronym[shortplural={O-DUs},longplural={Open Distributed Units}]{odu}{O-DU}{Open Distributed Unit}
\newacronym{ngn}{NGN}{Next-Generation Network}
\newacronym{mec}{MEC}{Multi-Access Edge Computing}
\newacronym{iot}{IoT}{Internet of Things}
\newacronym[shortplural={RANs},longplural={Radio Access Networks}]{ran}{RAN}{Radio Access Network}
\newacronym[shortplural={O-RANs},longplural={Open Radio Access Networks}]{oran}{O-RAN}{Open Radio Access Network}
\newacronym{3gpp}{3GPP}{3rd Generation Partnership Project}
\newacronym{qos}{QoS}{Quality of Service}
\newacronym{qoe}{QoE}{Quality of Experience}
\newacronym[shortplural={CNFs}, longplural={Cloud-Native Network Functions}]{cnf}{CNF}{Cloud-Native Network Function}
\newacronym[shortplural={RICs},longplural={RAN Intelligent Controllers}]{ric}{RIC}{RAN Intelligent Controller}
\newacronym{nfv}{NFV}{Network Function Virtualization}
\newacronym{cni}{CNI}{Container Network Interface}
\newacronym{rt}{RT}{Real Time}
\newacronym{nearrt}{Near-RT}{Near-Real Time}
\newacronym{nonrt}{Non-RT}{Non-Real Time}
\newacronym[shortplural={APIs}, longplural={Application Programming Interfaces}]{api}{API}{Application Programming Interface}
\newacronym{smo}{SMO}{Service Management and Orchestration}
\newacronym{tsn}{TSN}{Time Sensitive Networking}
\newacronym{sdl}{SDL}{Shared Data Layer}
\newacronym{slo}{SLO}{Service Level Objectives}
\newacronym{sla}{SLA}{Service Level Agreement}
\newacronym{sba}{SBA}{Service-based Architecture}
\newacronym{focom}{FOCOM}{Federated O-Cloud Orchestration \& Management}
\newacronym{fcaps}{FCAPS}{Fault, configuration, accounting, performance and security (FCAPS)}
\newacronym[shortplural={gNBs},longplural={5G next-generation Nodes B}]{gnb}{gNB}{5G next-generation Node B}
\newacronym[shortplural={UEs},longplural={User Equipments}]{ue}{UE}{User Equipment}
\newacronym{ml}{ML}{Machine Learning}
\newacronym[shortplural={NFs},longplural={Network Functions}]{nf}{NF}{Network Function}
\newacronym{5gc}{5GC}{5G Core}
\newacronym{epc}{EPC}{Evolved Packet Core}
\newacronym{cn}{CN}{Core Network}
\newacronym{enb}{eNB}{4G Evolved Node B}
\newacronym{lte}{LTE}{Long-Term Evolution}
\newacronym{ocloud}{O-Cloud}{Open Cloud}
\newacronym{rf}{RF}{Radio Frequency}
\newacronym{phy}{PHY}{Physical}
\newacronym{phylo}{PHY-low}{Physical-low}
\newacronym{phyhi}{PHY-high}{Physical-high}
\newacronym{mac}{MAC}{Medium Access Control}
\newacronym{rlc}{RLC}{Radio Link Control}
\newacronym{pdcp}{PDCP}{Packet Data Convergence Protocol}
\newacronym{sdap}{SDAP}{Service Data Adaptation Protocol}
\newacronym{rrc}{RRC}{Radio Resource Control}
\newacronym{bbu}{BBU}{Baseband Unit}
\newacronym[shortplural={KPMs},longplural={Key Performance Measurements}]{kpm}{KPM}{Key Performance Measurement}
\newacronym[shortplural={KPIs},longplural={Key Performance Indicators}]{kpi}{KPI}{Key Performance Indicator}
\newacronym{sinr}{SINR}{Signal to Interference-plus-Noise Ratio}
\newacronym[shortplural={SMs},longplural={Service Models}]{sm}{SM}{Service Model}
\newacronym[shortplural={E2SMs},longplural={E2 Service Models}]{e2sm}{E2SM}{E2 Service Model}
\newacronym{e2smkpm}{E2SM-KPM}{E2 Service Model for Key Performance Measurements}
\newacronym{e2smrc}{E2SM-RC}{E2 Service Model for RAN Control}
\newacronym{e2smni}{E2SM-NI}{E2 Service Model for Network Interfaces}
\newacronym{e2smccc}{E2SM-CCC}{E2 Service Model for Cell Configuration and Control}
\newacronym{rai}{RAI}{Radio Analytics Information}
\newacronym[shortplural={DRBs},longplural={Data Radio Bearers}]{drb}{DRB}{Data Radio Bearer}
\newacronym[shortplural={SRBs},longplural={Signal Radio Bearers}]{srb}{SRB}{Signal Radio Bearer}
\newacronym{fapi}{FAPI}{5G Femto Application Platform Interface}
\newacronym{nfapi}{nFAPI}{network Functional Application Platform Interface}
\newacronym{sa}{SA}{Stand Alone}
\newacronym{nr}{nr}{New Radio}
\newacronym{devops}{DevOps}{Development and Operations}
\newacronym{ci}{CI}{Continuous Integration}
\newacronym{cd}{CD}{Continuous Deployment}
\newacronym{scf}{SCF}{Small Cell Forum}
\newacronym{oransc}{OSC}{O-RAN Software Community}

\newacronym{rmr}{RMR}{RIC Message Router}
\newacronym{e2m}{E2M}{E2 Manager}
\newacronym{e2t}{E2T}{E2 Termination}
\newacronym{subm}{SubM}{Subscription Manager}
\newacronym{a1m}{A1M}{A1 Mediator}
\newacronym{rm}{RM}{Routing Manager}
\newacronym{xappm}{xAppM}{xApp Manager}
\newacronym{iac}{IaC}{Infrastructure as Code}
\newacronym{dbaas}{DBaaS}{DataBase as a Service}
\newacronym{sdn}{SDN}{Software Defined Networks}
\newacronym{crd}{CRD}{Custom Defined Resources}
\newacronym{alm}{AlM}{Alarm Manager}
\newacronym{vespam}{VESPAM}{Virtual Event Streaming Prometheus Adapter Manager}
\newacronym{mno}{MNO}{Mobile Network Operator}
\newacronym{crud}{CRUD}{Create, Read, Update, and Delete}
\newacronym{airan}{AI-RAN}{Artificial Intelligence Radio Access Network}
\newacronym{oai}{OAI}{OpenAirInterface}

\newacronym{mtls}{mTLS}{mutual TLS}
\newacronym{xds}{xDS}{xDiscovery Service}
\newacronym{lds}{LDS}{Listener Discovery Service}
\newacronym{cds}{CDS}{Cluster Discovery Service}
\newacronym{rds}{RDS}{Route Discovery Service}
\newacronym{eds}{EDS}{Endpoint Discovery Service}
\newacronym{sds}{SDS}{Secret Discovery Service}
\newacronym{gui}{GUI}{Graphic User Interface}
\newacronym{cli}{CLI}{Command User Interface}
\newacronym{cicd}{CI/CD}{Continuous Integration/Continuous Deployment}
\newacronym{mcp}{MCP}{Model Context Protocol}
\newacronym{oursol}{MANATEE}{Mesh Architecture for Radio Access Network Automation and TEsting Ecosystem}
\newacronym{ebpf}{eBPF}{extended Berkeley Packet Filter}
\newacronym{llm}{LLM}{Large Language Model}
\newacronym{sdr}{SDR}{Software-defined Radio}
\newacronym{cncf}{CNCF}{Cloud Native Computing Foundation}
\newacronym{rc}{RC}{RAN Control}
\newacronym{prb}{PRB}{Physical Resource Block}
\newacronym{plmn}{PLMN}{Public Land Mobile Network}

\newcommand{\oursol}[0]{\gls{oursol}\xspace}
\newcommand{\nearrtric}{\gls{nearrt} \gls{ric}\xspace}
\newcommand{\nonrtric}{\gls{nonrt} \gls{ric}\xspace}

\usepackage{tikzpagenodes,etoolbox}
\usetikzlibrary{calc}
\usepackage[contents={}]{background}
\AddEverypageHook{%
\ifnumequal{\thepage}{1}{%
    \tikz[remember picture,overlay]{%
        \node[draw,
        minimum width=1.07\textwidth,
        text width=1.07\textwidth,
        font=\footnotesize
        ]
        at ($(current page header area) - (0,-11pt)$)
        {%
        This work has been submitted to the IEEE for possible publication. Copyright may be transferred without notice, after which this version may no longer be accessible.
        };
    }%
}{}%end ifnumequal
}

\begin{document}

\title{MANATEE: A DevOps Platform for xApp Lifecycle Management and Testing in Open RAN}

\author{Sofia Montebugnoli, Leonardo Bonati, Andrea Sabbioni,  Luca Foschini, Paolo Bellavista, Salvatore D'Oro, Michele Polese, Tommaso Melodia

\thanks{S. Montebugnoli, A. Sabbioni, L. Foschini, and  P. Bellavista are with the Department of Computer Science and Engineering, University of Bologna, Bologna, 40126, Italy. L. Bonati, S. D'Oro, M. Polese, and T. Melodia are with the Institute for the Wireless Internet of Things, Northeastern University, Boston, MA, 02115, USA. E-mail: \{sofia.montebugnoli3, andrea.sabbioni5, luca.foschini, paolo.bellavista\}@unibo.it, \{l.bonati, s.doro, m.polese, t.melodia\}@northeastern.edu.}
}

\markboth{Journal of \LaTeX\ Class Files,~Vol.~XX, No.~X, January~2026}%
{Montebugnoli \MakeLowercase{\textit{et al.}}: Bare Demo of IEEEtran.cls for IEEE Journals}

\maketitle

\begin{abstract}
The shift to disaggregated 5G architectures introduces unprecedented flexibility but also significant complexity in Beyond 5G \glspl{ran}. Open \gls{ran} enables programmability through xApps, yet deploying and validating these applications is critical given the nature of the systems they aim to control. Current Open RAN ecosystems lack robust lifecycle management of xApps that enable automated testing, seamless migration, and production-grade observability, resulting in slow, error-prone xApp delivery.
To address these issues, DevOps practices can streamline the xApp lifecycle by integrating \gls{cicd} pipelines with advanced traffic management and monitoring, such as leveraging service mesh technologies to enable progressive deployment strategies (e.g., canary releases and A/B testing) to ensure fine-grained observability and resilience. The solution presented in this article, MANATEE (\textbf{M}esh \textbf{A}rchitecture for Radio Access \textbf{N}etwork \textbf{A}utomation and \textbf{TE}sting \textbf{E}cosystems), is the first platform that combines these principles to simplify xApp delivery into production, accelerate innovation, and guarantee performance across heterogeneous O-RAN environments. We prototyped MANATEE on a Kubernetes cluster integrated with the O-RAN Software Community \acrlong{nearrt} \gls{ric}, as well as with service mesh technologies, to facilitate testing of xApps across simulated, emulated, and real testbed environments.
Our experimental results demonstrate that service mesh integration introduces minimal overhead (below 1\:ms latency), while enabling reliable canary deployments with fine-grained traffic control and conflict-free A/B testing through circuit-breaking mechanisms. 
\end{abstract}

\IEEEpeerreviewmaketitle

\glsresetall

%%%%%%%%%%%%%%%%%%%%%%%%%%%%%%%%%%%%%%%%%%%%%%%%%%%%%%%%%%%%%%%%%%%%%%%%%%%%%%%%%%%%%%%%%%%%%%%%%%%%%%%%%%%%%%%%%%
%%%%%%%%%%%%%%%%%%%%%%%%%%%%%%%%%%%%%%%%%%%%%%%%%%%%%%%%%%%%%%%%%%%%%%%%%%%%%%%%%%%%%%%%%%%%%%%%%%%%%%%%%%%%%%%%%%
%INTRODUCTION
%%%%%%%%%%%%%%%%%%%%%%%%%%%%%%%%%%%%%%%%%%%%%%%%%%%%%%%%%%%%%%%%%%%%%%%%%%%%%%%%%%%%%%%%%%%%%%%%%%%%%%%%%%%%%%%%%%
%%%%%%%%%%%%%%%%%%%%%%%%%%%%%%%%%%%%%%%%%%%%%%%%%%%%%%%%%%%%%%%%%%%%%%%%%%%%%%%%%%%%%%%%%%%%%%%%%%%%%%%%%%%%%%%%%%
\section{Introduction}

The evolution of mobile networks has led to increasingly complex \gls{ran} architectures aimed at meeting growing demands for capacity, coverage, and service quality. Traditional \gls{ran} systems rely on proprietary and vendor-specific solutions with tightly coupled hardware and software components, forming closed ecosystems that severely limit flexibility and innovation~\cite{understandingoran2023polese}. This monolithic design has resulted in vendor lock-in, high deployment costs, and slow innovation cycles.
To overcome these limitations, Open \gls{ran} and, specifically, its O-RAN embodiment, introduce a paradigm shift toward openness and disaggregation. By defining open and standardized interfaces between the cellular network components, O-RAN enables the development of intelligent, virtualized, and interoperable solutions to dynamically reconfigure network parameters at run-time~\cite{wg1}. Central to this architecture is the \nearrtric, which enables fine-grained control of \gls{ran} behavior through third-party applications known as xApps, allowing operators to instantiate dynamic optimization strategies tailored to specific deployment scenarios~\cite{wg1}.

This transformation toward a flexible and programmable network introduces new challenges for developers of \gls{ran} components, who must deliver high-quality, robust, and high-performing software in increasingly complex and diverse environments. Unlike traditional software systems, xApps operate as closed-loop controllers that directly influence live cellular networks, issuing control commands to \glspl{gnb}, adjusting scheduling policies, and reconfiguring radio parameters in real time~\cite{wg1}. This tight coupling with production infrastructure means that software flaws, misconfigurations, or performance regressions can immediately degrade network service for end users. Consequently, xApp lifecycle management demands rigorous testing and deployment practices that account for this operational criticality.

Despite these requirements, applying DevOps practices to automate O-RAN lifecycle management remains challenging. The \gls{oransc} \nearrtric lacks native support for traffic splitting and progressive rollouts, requiring manual configuration of routing rules for each xApp version. Similarly, transitioning an xApp from a simulated E2 node to a real \gls{gnb} requires reconfiguring connection endpoints, re-registering with the Application Manager, and manually validating health probes, steps that are error-prone and difficult to automate without dedicated tooling. Moreover, current platforms provide rigid, tightly coupled components that are hard to monitor, manage, and reconfigure~\cite{tutorial5g2025, maxenti2025autoranautomatedzerotouchopen}. The absence of real-world testing environments and limited observability further hinders reliable xApp deployment at scale~\cite{tutorial5g2025, fromzerotohero2025}.

\begin{figure}[!ht]
\centering
\includegraphics[width=\columnwidth]{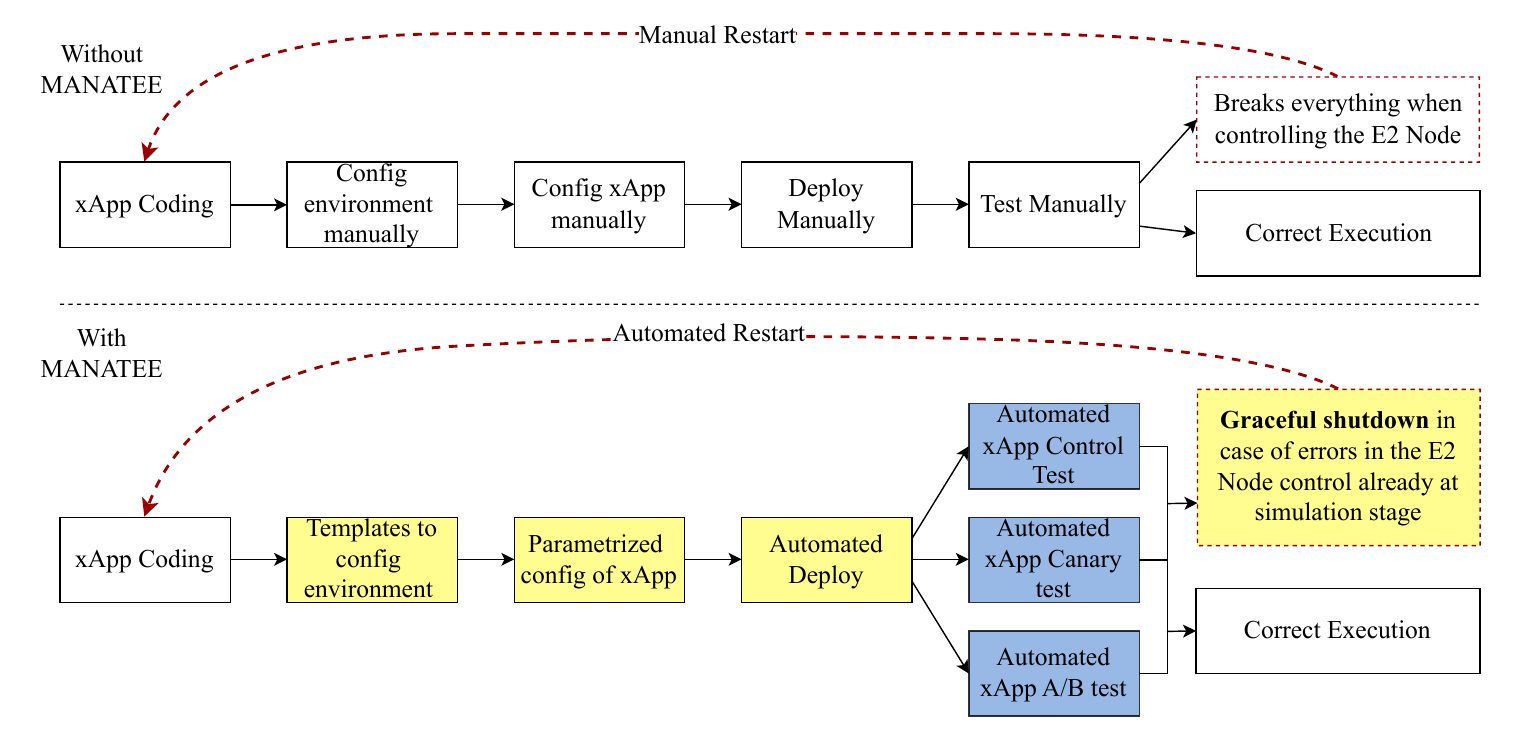}
\caption{Comparison of a classic flow with and without MANATEE.}
\label{fig:manatee-intro}
\end{figure}

To address these limitations, we introduce \oursol, a DevOps-enabled platform that manages the entire xApp lifecycle from early testing to in-production quality assurance. By embedding \gls{cicd} practices into the \gls{ran} ecosystem, \oursol enables faster and safer deployments, progressive rollout strategies, and continuous observability. The platform bridges the gap between simulated, emulated, and real testbeds, ensuring reproducible deployments and enabling stress and resilience testing under realistic network conditions. While our prototype focuses on xApps, the underlying approach applies broadly to any application controlling cellular network infrastructure, including network functions and orchestration components.

Figure~\ref{fig:manatee-intro} illustrates the differences between the non-automated approach and \oursol. Our contributions are as follows:
\begin{itemize}
    \item[--] A modular architecture that decouples xApp deployment from environment-specific configurations, enabling seamless migration across simulation, emulation, and production testbeds.
    \item[--] Service mesh integration for fine-grained traffic management, supporting canary deployments, A/B testing, and traffic shifting without xApp modification.
    \item[--] Comprehensive observability through distributed tracing, metrics collection, and centralized logging for auditing xApp behavior and performance.
    \item [--] Automated xApp lifecycle management that orchestrates version deployments, progressive traffic shifts, and health-driven rollbacks without manual intervention, enabling safe testing-in-production workflows at scale.
\end{itemize}

Different from what is available most often in the related literature, we have performed a careful in-the-field validation of our platform in a real multi-environment testbed comprising simulated E2 nodes, emulated \glspl{gnb}, and production-grade infrastructure. Our evaluation covers three scenarios reflecting real-world challenges: xApp migration across environments, canary deployments with progressive traffic shifting, and A/B testing for direct performance comparison. Our solution is publicly available as open source to foster reproducibility and further research~\footnote{https://github.com/wineslab}.

Our experiments demonstrate that service mesh integration introduces negligible overhead, with less than $10$\% impact on Kubernetes operations and sub-millisecond latency for xApp communication. Canary deployments maintain zero errors under load up to $600$\:msg/s, whereas A/B testing achieves conflict-free operation via circuit-breaking mechanisms with transition delays of only $6$\:s.

The remainder of this paper is structured as follows. In Sec.~\ref{sec:background}, we provide background information on the main concepts used throughout the paper. In Sec.~\ref{sec:relatedworks}, we review existing literature approaches related to \oursol. In Sec.~\ref{sec:architecture}, we present our proposed architecture, while we detail \oursol lifecycle management and in Sec.~\ref{sec:testingflows}.
In Sec.~\ref{sec:implementation}, we describe the \oursol prototype.
In Sec.~\ref{sec:experimentalresults}, we present our experimental results comparing various service mesh approaches, and showing the behavior of various testing-in-production practices. Finally, in Sec.~\ref{sec:conclusions}, we summarize our findings and outline possible extensions of our work.

%%%%%%%%%%%%%%%%%%%%%%%%%%%%%%%%%%%%%%%%%%%%%%%%%%%%%%%%%%%%%%%%%%%%%%%%%%%%%%%%%%%%%%%%%%%%%%%%%%%%%%%%%%%%%%%%%%
%%%%%%%%%%%%%%%%%%%%%%%%%%%%%%%%%%%%%%%%%%%%%%%%%%%%%%%%%%%%%%%%%%%%%%%%%%%%%%%%%%%%%%%%%%%%%%%%%%%%%%%%%%%%%%%%%%
%BACKGROUND
%%%%%%%%%%%%%%%%%%%%%%%%%%%%%%%%%%%%%%%%%%%%%%%%%%%%%%%%%%%%%%%%%%%%%%%%%%%%%%%%%%%%%%%%%%%%%%%%%%%%%%%%%%%%%%%%%%
%%%%%%%%%%%%%%%%%%%%%%%%%%%%%%%%%%%%%%%%%%%%%%%%%%%%%%%%%%%%%%%%%%%%%%%%%%%%%%%%%%%%%%%%%%%%%%%%%%%%%%%%%%%%%%%%%%
\begin{figure}[!ht]
\centering
\includegraphics[width=\columnwidth]{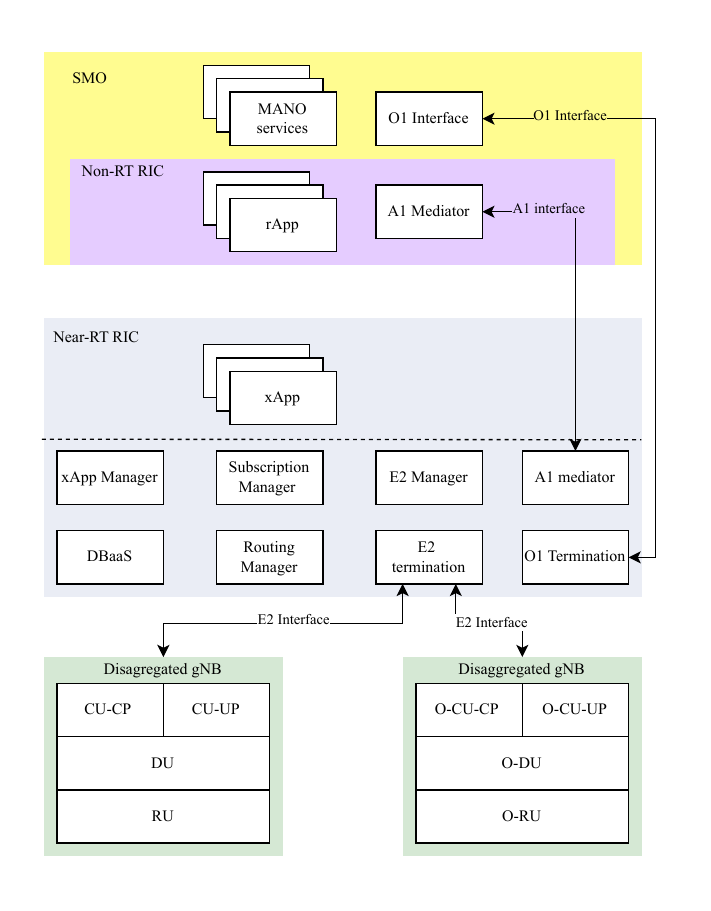}
\caption{O-RAN architecture with focus on the \acrshort{nearrt} \acrshort{ric}.}
\label{fig:oransc-helm}
\end{figure}

\section{Background}\label{sec:background}
This section provides background on the Open \gls{ran} architecture, DevOps methodologies, and service mesh technologies that enable the automation and orchestration presented in this work.

\subsection{Open RAN Architecture}

The Open \gls{ran} architecture, shown in Fig.~\ref{fig:oransc-helm}, promotes a vendor-neutral, flexible, and programmable mobile network infrastructure~\cite{wg1}. Building upon disaggregated 7.2 split \gls{gnb} components (\gls{ru}, \gls{du}, \gls{cu}), open and programmable interfaces and a \gls{smo}, the O-RAN Alliance defines additional intelligent control elements, such as the \nearrtric, \nonrtric, and the O-Cloud.

The \nearrtric enables near-real-time optimization and control of \gls{cu} and \gls{du} components on 10 ms to 1 s timescales, with xApps serving as the primary customization point for developers to implement \gls{ran} control logic. Similarly, the \nonrtric operates on longer timescales ($>1$\:s) for non-real-time control and optimization of \glspl{gnb}, offering rApps as customization points for non-real-time control algorithms. The \gls{smo} serves as the management and orchestration layer, responsible for both network elements and the platform infrastructure itself. These components communicate through standardized interfaces: A1 between \nearrtric and \nonrtric, and E2 between \nearrtric and disaggregated \gls{gnb}/\gls{enb}~\cite{wg1}.

Central to realizing automation in Open \gls{ran} is the availability of flexible testing and validation platforms that support the entire development lifecycle, from simulation to emulation to production deployment. Such platforms need to emulate E2 node behavior and network dynamics, enabling developers to verify conformance, assess performance, and validate fault tolerance through automated testing workflows before production integration. The O-RAN Test and Integration Focus Group (TIFG) coordinates these efforts by defining comprehensive testing plans and maintaining reference platforms to ensure interoperability across vendors and components~\cite{TIFG}. However, bridging the gap between reference implementations and production-ready platforms remains a critical challenge for deploying robust xApps in open and disaggregated \gls{ran} environments.

\subsection{DevOps for Open RAN}

DevOps is a paradigm that unifies software development and operations into a continuous lifecycle management platform for IT applications. DevOps practices are closely tied to containerization and microservice architectures. A microservice application, decomposed into loosely coupled services, enables independent deployment and scaling that eases its DevOps lifecycle. Without the pretense of being exhaustive, let us remind that, in DevOps approaches, lifecycle management is structured into eight phases, namely Plan, Code, Build, Test, Release, Deploy, Operate, and Monitor, each underpinned by specialized tools and automation frameworks that enable rapid and reliable software delivery~\cite{kim2016devops}. The benefits of DevOps are manifold and include accelerated time-to-market, reduced lead time to release updates, lower change failure rates, lower mean time to recovery, and improved cross-team collaboration and innovation. 

Open \gls{ran} emphasizes the adoption of cloud-native principles and DevOps practices as fundamental enablers for automation in \gls{ran} deployments \cite{intelligentoranbeyond5g2024}. DevOps methodologies are pivotal for accelerating \gls{cnf} deployment, streamlining lifecycle management, and enabling continuous integration and delivery of softwarized \gls{ran} components~\cite{wg6}. This transformation requires automated testing pipelines, progressive deployment strategies, and comprehensive observability frameworks that traditionally characterize modern software engineering but remain underutilized in \gls{ran} environments.

Focusing on testing, DevOps follows a pyramidal model with extensive unit testing across the codebase, typically triggered on each commit. Integration and \gls{api} tests validate component interactions during release cycles, and component tests validate individual components or services in isolation, with external dependencies replaced by test doubles or mocks. In contrast, E2E tests verify complete user workflows across the whole system. In-production testing extends beyond traditional pre-production validation by deploying and evaluating new xApp releases progressively: first in testing environments with simulated E2 terminations and \gls{ran} deployments using emulated \gls{rf} environments (such as ORANInABox or Colosseum \cite{colosseum2021,xdevsmferaudo2024}); and then advancing to production environments with Open \gls{ran} components operating in live network conditions.

\subsection{Service Mesh Integration}

An important approach that plays a pivotal role in enabling DevOps in complex scenarios with many entangled services is service mesh.
A service mesh is a dedicated infrastructure layer that provides secure, fast, and reliable service-to-service communication for distributed microservice-based applications~\cite{cncfmesh}. It consists of two primary components. 
The data plane comprises all the sidecar proxies that route inbound and outbound traffic for each microservice, while the control plane provides the configuration of transparent traffic management, security, and observability for the sidecar proxies without changes to application code \cite{cncfmesh}. 
By decoupling communication management from business logic, service meshes enable fine-grained control over service-to-service communication~\cite{lookatservicemesh}. For example, fine-grained control and observability are crucial for next-generation network workloads that require near-real-time control and high reliability \cite{3gpp2023ts23501}.

Service meshes are essential for DevOps in Open \gls{ran} because they decouple communication management from application logic and directly embed advanced traffic control, security, and observability into \gls{ci}/\gls{cd} pipelines \cite{challengesopportunitiesservicemesh, soldanisauron2023}. These capabilities enable progressive deployment strategies such as Canary Deployment, which involves routing a small fraction of production traffic to a new \gls{cnf} version and gradually increasing it based on health metrics, minimizing risk during rollouts. Similarly, A/B testing enables controlled experiments to compare two or more variants of an application and measure predefined metrics, enabling data-driven validation of features before full release \cite{10719389, 10.1007/978-3-030-33702-5_12, 10.1145/3194760.3194763}. These capabilities can be integrated into DevOps pipelines to declaratively manage mesh configurations and automate traffic-shifting policies as part of \gls{cicd} workflows \cite{10719389, soldanisauron2023}, ultimately reducing time-to-market and ensuring robust, production-ready deployments in 5G and beyond environments \cite{servicemeshreadiness5g2023, duongservicemeshcore5g2023}.

\begin{figure*}[!t]
    \centering
    \begin{subfigure}[t]{0.32\textwidth}
        \includegraphics[width=1.2\textwidth]{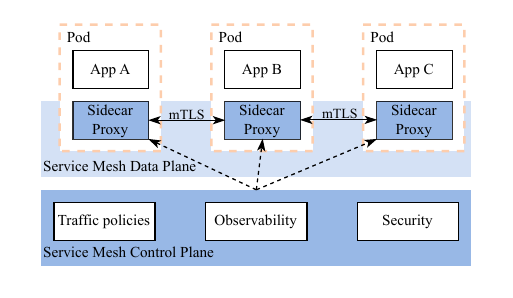}
        \caption{Service mesh architecture}
        \label{subfig:servicemesh}
    \end{subfigure}
    \hfill
    \begin{subfigure}[t]{0.32\textwidth}
        \includegraphics[width=1.2\textwidth]{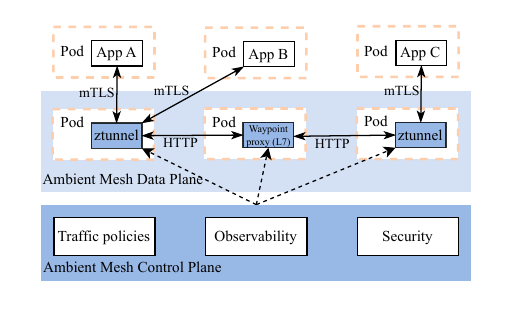}
        \caption{Ambient mesh proxyless architecture}
        \label{subfig:ambientmesh}
    \end{subfigure}
    \hfill
    \begin{subfigure}[t]{0.32\textwidth}
        \includegraphics[width=1.1\textwidth]{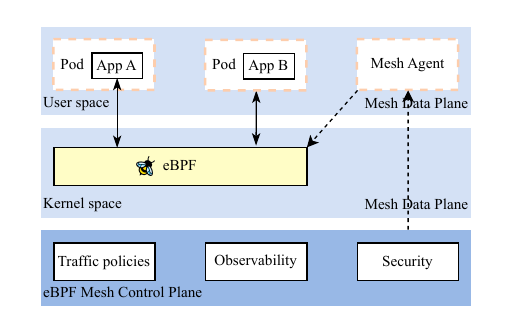}
        \caption{eBPF mesh proxyless architecture}
        \label{subfig:ebpfmesh}
    \end{subfigure}
    \caption{Architecture comparison of different types of mesh technologies.}
    \label{fig:emrt-comparison}
\end{figure*}

Service mesh architectures can vary based on the deployment approach for the data plane \cite{ambient, cilium, istio}.
Conventional service mesh implementations, shown in Fig.~\ref{subfig:servicemesh},  utilize per-service sidecar proxies to provide traffic control, telemetry collection, and security policy enforcement \cite{istio}. Although this architecture offers complete feature coverage, it incurs substantial performance overhead \cite{zhu2022dissecting}, driving innovation toward optimized proxy designs \cite{servicemesh_ebpf, servicemeshmeta} and sidecarless deployment models \cite{ambient, cilium}. These sidecarless approaches retain the core service mesh abstractions (control plane and data plane separation) while implementing the data plane differently to reduce resource overhead.

Figure~\ref{subfig:ambientmesh} depicts the Ambient mesh architecture, a sidecar-less service mesh approach that reduces complexity and resource overhead by decoupling data-plane functions~\cite{ambient}. Rather than injecting sidecar proxies into each application pod, Ambient Mesh employs a layered architecture. At its foundation is the ztunnel, a lightweight, shared agent that handles secure routing and zero-trust authentication (Layer 4) at the node level. When more advanced Layer 7 features are needed, such as traffic shaping or observability, waypoint proxies, and Envoy-based pods can be selectively deployed at the namespace level. This approach improves scalability, simplifies operational management, and allows for gradual adoption without disrupting running workloads \cite{ambient}. Furthermore, it preserves interoperability with traditional sidecar-based deployments.

More recently, \gls{ebpf}-based meshes, such as Cilium, also adopt a sidecarless model but rely on a fundamentally different mechanism compared to Ambient mesh \cite{cilium, servicemesh_ebpf}. As shown in Fig.~\ref{subfig:ebpfmesh}, Cilium connects services via native TCP sockets and performs data-plane operations using in-kernel \gls{ebpf} code. This allows \gls{ebpf} mesh to enforce security policies, perform traffic routing, and enable observability directly within the Linux kernel, resulting in lower latency and improved performance \cite{servicemesh_ebpf}. Unlike approaches that require application-level proxies, \gls{ebpf} mesh avoids the need for sidecars entirely and integrates tightly with the Kubernetes networking substrate. It can also be integrated with higher-level service mesh abstractions as needed, providing a lightweight, high-performance alternative for secure service-to-service communication.

%%%%%%%%%%%%%%%%%%%%%%%%%%%%%%%%%%%%%%%%%%%%%%%%%%%%%%%%%%%%%%%%%%%%%%%%%%%%%%%%%%%%%%%%%%%%%%%%%%%%%%%%%%%%%%%%%%
%%%%%%%%%%%%%%%%%%%%%%%%%%%%%%%%%%%%%%%%%%%%%%%%%%%%%%%%%%%%%%%%%%%%%%%%%%%%%%%%%%%%%%%%%%%%%%%%%%%%%%%%%%%%%%%%%%
%RELATED WORKS
%%%%%%%%%%%%%%%%%%%%%%%%%%%%%%%%%%%%%%%%%%%%%%%%%%%%%%%%%%%%%%%%%%%%%%%%%%%%%%%%%%%%%%%%%%%%%%%%%%%%%%%%%%%%%%%%%%
%%%%%%%%%%%%%%%%%%%%%%%%%%%%%%%%%%%%%%%%%%%%%%%%%%%%%%%%%%%%%%%%%%%%%%%%%%%%%%%%%%%%%%%%%%%%%%%%%%%%%%%%%%%%%%%%%%

\section{Related Work}\label{sec:relatedworks}

% \hl{LB: this section only surveys few works and is mostly self-referential. It would be good to have around 10 more works from other groups}
% {\color{red} ANDREA: ok summary e critica, sono molto lunghi andrebbero aggregati o messo un aggancio a manatee alla fine di ognuno,ad es in questo caso: "Our solution, manatee overcome this limitations by offering an end-to-end solution for xapp testing supporting developer to deliver production-ready xapps" LB: agree, I'm giving it a pass}.
% \hlcyan{SM: I rewrote the section, reducing it and adding more papers as suggested by Leo during meetings, now we cite 13 works. LB: good to me, thank you}

While Open \gls{ran} has enabled unprecedented programmability and intelligence in cellular networks, comprehensive frameworks for testing and deploying xApps through their complete lifecycle remain largely unexplored. This section analyzes existing work in Open \gls{ran} automation, xApp development, and service mesh integration, highlighting how \oursol addresses critical gaps in xApp lifecycle management.
 
AutoRAN~\cite{maxenti2025autoranautomatedzerotouchopen} and 5G-CT~\cite{5gct2025} respectively present automated Open \gls{ran} deployment and testing frameworks based on Red Hat OpenShift. The former leverages cloud-native principles and \gls{cicd} pipelines to deploy cellular network components in a zero-touch and intent-driven manner. The latter focuses on the automated deployment and testing of such components leveraging GitOps workflows.
Similarly, OrchestRAN~\cite{orchestran2022} presents an orchestration framework for automating network intelligence in Open \gls{ran}, while the work in~\cite{9647628} proposes automated \gls{cicd} of network applications over 5G infrastructure. 
Although these works advance automation and orchestration in 5G and O-RAN environments, they do not address end-to-end xApp lifecycle management, as instead \oursol does by integrating \gls{cicd} pipelines with service mesh capabilities.

The work in~\cite{9376232} presents the \gls{oransc} \nearrtric platform design, including \gls{rmr} for inter-xApp messaging. While \gls{rmr} implements the publish-subscribe communication paradigm, its design precedes modern service mesh architectures and lacks advanced traffic management functionalities. \oursol augments the \nearrtric with service mesh capabilities, enabling fine-grained traffic control.
% without requiring modifications to the xApps.
Studies in~\cite{10329915} and~\cite{fromzerotohero2025} analyze xApp design challenges, identifying obstacles including complex deployment files and limited version comparability. \oursol directly addresses these challenges by automating deployments, enabling A/B testing for version comparison, and providing observability for performance evaluation.
The work in~\cite{9681936} implements an O-RAN AI/ML workflow for personalized network optimization but does not address the DevOps challenges related to deploying such models in production. \oursol, instead, supports MLOps-style workflows by enabling safe canary deployments and rollback mechanisms for AI-driven xApps.
Testbeds such as CCI xG~\cite{10858200} and enterprise-scale platforms~\cite{10.1145/3570361.3615745} provide valuable experimental infrastructure but lack integrated DevOps tooling. \oursol complements such testbeds by providing the automation layer needed to transition xApps from experimental validation to production environments.
 
The first comprehensive analysis of service mesh technologies for deploying xApps in the \nearrtric is presented in \cite{10901151}, which evaluates their impact on communication, observability, and traffic management.
Similarly, the work in~\cite{10327727} investigates service mesh readiness for 5G, validating its applicability in telecom environments but not addressing xApp deployment.
While representing important first steps, these works do not integrate service mesh with \gls{cicd} pipelines. \oursol builds upon these findings by leveraging service mesh for automated canary deployments and A/B testing within DevOps workflows, and extends its adoption to the \nearrtric for xApp traffic management. 
Finally, \cite{fi17080372} proposes zero trust architectures for O-RAN using Kubernetes and Istio, focusing on security rather than on lifecycle management. \oursol leverages similar technologies but it focuses on enabling safe in-production testing.

To the best of our knowledge, \oursol is the first platform comprehensively addressing xApp lifecycle management in production environments. By combining \gls{cicd} pipelines with service mesh capabilities tailored to the \nearrtric, \oursol enables: (i)~automated xApp registration and deployment; (ii)~in-production testing techniques such as canary deployment and A/B testing; (iii)~comprehensive observability for live version comparison; and (iv)~seamless transitions between simulated, emulated, and real testbed environments.

%%%%%%%%%%%%%%%%%%%%%%%%%%%%%%%%%%%%%%%%%%%%%%%%%%%%%%%%%%%%%%%%%%%%%%%%%%%%%%%%%%%%%%%%%%%%%%%%%%%%%%%%%%%%%%%%%%
%%%%%%%%%%%%%%%%%%%%%%%%%%%%%%%%%%%%%%%%%%%%%%%%%%%%%%%%%%%%%%%%%%%%%%%%%%%%%%%%%%%%%%%%%%%%%%%%%%%%%%%%%%%%%%%%%%
%MANATEE
%%%%%%%%%%%%%%%%%%%%%%%%%%%%%%%%%%%%%%%%%%%%%%%%%%%%%%%%%%%%%%%%%%%%%%%%%%%%%%%%%%%%%%%%%%%%%%%%%%%%%%%%%%%%%%%%%%
%%%%%%%%%%%%%%%%%%%%%%%%%%%%%%%%%%%%%%%%%%%%%%%%%%%%%%%%%%%%%%%%%%%%%%%%%%%%%%%%%%%%%%%%%%%%%%%%%%%%%%%%%%%%%%%%%%

\begin{figure*}[!ht]
\centering
\includegraphics[width=\textwidth]{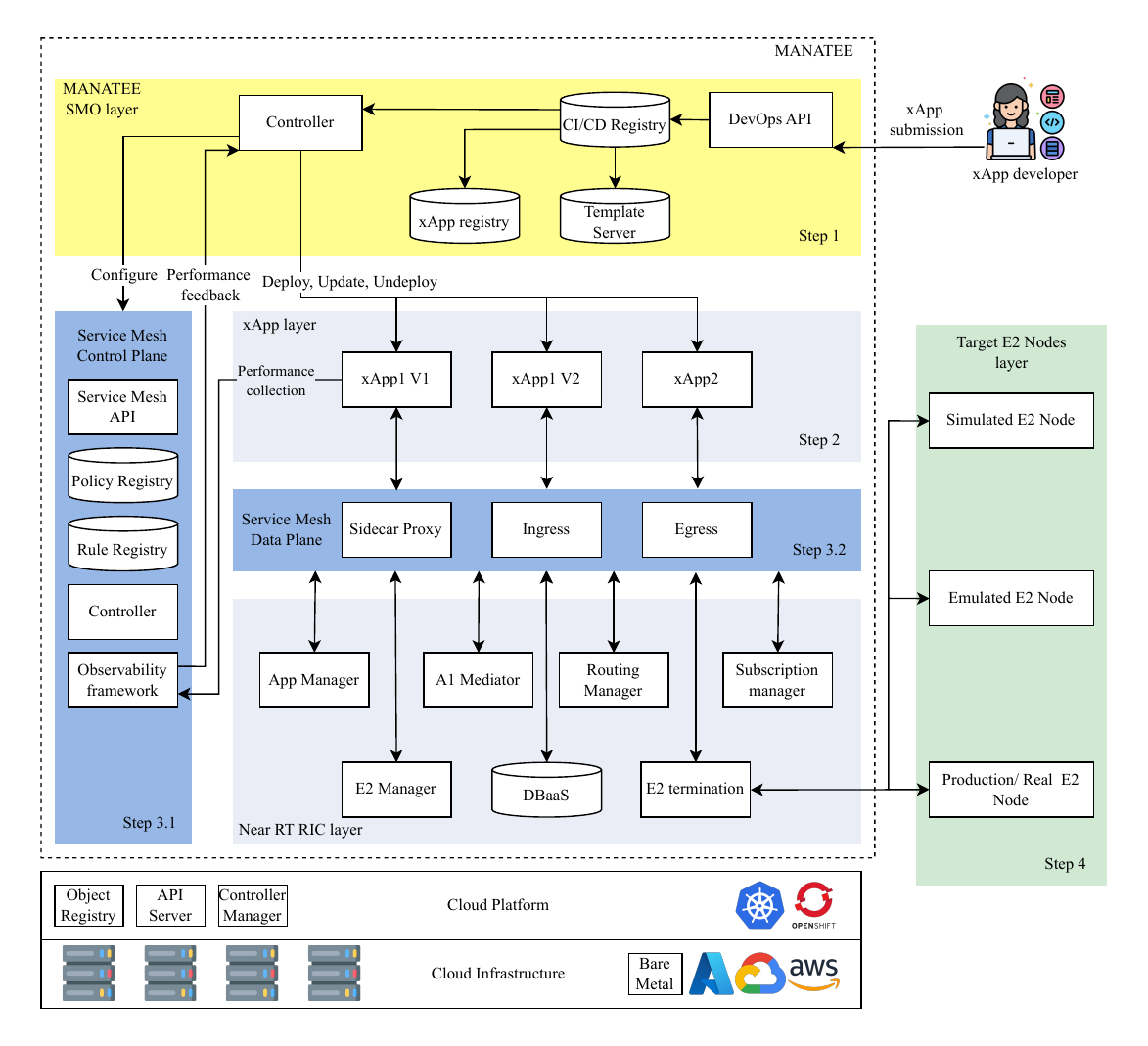}
\caption{\oursol architecture.}
\label{fig:manatee}
\end{figure*}

\begin{figure}[!ht]
\centering
\includegraphics[width=\columnwidth]{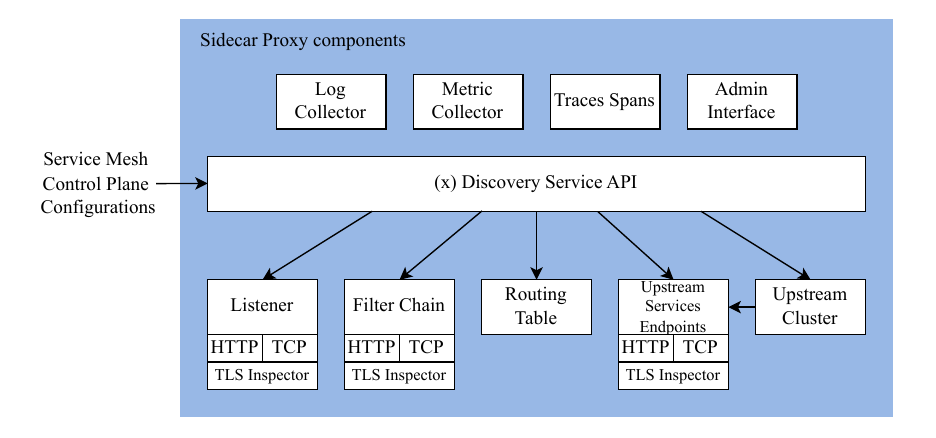}
\caption{Sidecar proxy architecture.}
\label{fig:sidecar-proxy}
\end{figure}

\section{MANATEE Architecture}
\label{sec:architecture}

Figure~\ref{fig:manatee} shows the overall \oursol architecture and its integration with the \gls{oransc} \nearrtric. It comprises eight main components: the \emph{\oursol \gls{smo} layer}, the \emph{Service Mesh Control Plane}, the \emph{Service Mesh Data Plane}, the \emph{xApp layer}, the \emph{\nearrtric layer}, the \emph{Cloud Platform}, the \emph{Cloud Infrastructure}, and the \emph{Target E2 nodes layer}.

The \emph{\oursol \gls{smo} layer} coordinates all \gls{cicd} workflows within \oursol, including five main blocks (shown in yellow in Fig.~\ref{fig:manatee}): the DevOps API, the Template Server, the \gls{cicd} Registry, the xApp Registry, and the Controller.
The DevOps API enables xApp developers to interact with \gls{cicd} pipelines, including defining, executing, and deleting \gls{oursol} in production testing workflows.
The Template Server stores all the parametrized workflow definitions. These templates contain placeholders that are instantiated at runtime with specific values, such as target E2 node identifiers, traffic increment percentages and time intervals for Canary deployments, duration and success criteria for A/B tests, and xApp version tags. This parameterization enables workflow reuse across different deployment scenarios without modifying the underlying pipeline logic.
The \gls{cicd} Registry stores pipeline definitions implementing testing-in-production strategies of \oursol \emph{Migration of xApp control over E2 nodes}, \emph{Canary Deployment of xApps}, \emph{A/B Testing of xApps}.
The xApp Registry collects all xApp definitions and parameters to make them available for deployment from the \gls{cicd} pipelines. 
The Controller reconciles the workflow state with the defined state and pulls monitoring data from the Observability framework deployed in the service mesh control plane.

The \emph{Service Mesh Control Plane}, shown as the vertical dark blue box in Fig.~\ref{fig:manatee}, provides high-level definitions of all configurations targeting the service mesh. In particular, it handles traffic management by discovering services through Kubernetes registries and using \gls{xds} servers to push configurations to the sidecar proxies via their discovery \gls{api}s (\gls{lds}, \gls{rds}, \gls{cds}, \gls{eds}). These configurations translate high-level routing rules into instructions executed by proxies. On the security side, it operates a Certification Authority Server that issues X.509 certificates, automatically rotates them, and securely delivers credentials through \gls{sds}. It also manages configurations by watching the Kubernetes \glspl{api}, validating settings before they are applied. The control plane maintains an in-memory cache of configurations and reports status back to Kubernetes. A dashboard offers visibility into the mesh topology and traffic flow.

The \emph{Service Mesh Data Plane}, shown as a horizontal dark blue box at the center of Fig.~\ref{fig:manatee}, consists of sidecar proxies, ingress gateways, and egress gateways. Each \nearrtric component and xApp receives a dedicated sidecar proxy that performs traffic routing, load balancing, \gls{mtls} encryption/decryption, circuit breaking, retry logic, telemetry collection (metrics, logs, traces), and authentication and authorization requests. Internally, as shown in Fig.~\ref{fig:sidecar-proxy}, sidecars use \gls{xds} \gls{api}s to dynamically configure components from the control plane, including listeners, filter chains, routing tables, upstream endpoints, and cluster definitions.

The \emph{Observability Framework} component scrapes metrics exposed by sidecar proxies and other components, storing them for querying and alerting. A customized dashboard visualizes these metrics, enabling real-time monitoring of xApp performance, service mesh health, and system-wide telemetry.

\oursol leverages either Kubernetes or OpenShift as \emph{cloud platforms}, while the underlying \emph{cloud infrastructure} can be deployed on either bare-metal or cloud-provider environments.
The \nearrtric connects to the \emph{E2 nodes layer} through the E2 termination component. The \emph{E2 node layer} could contain simulated, emulated, or real deployments. The specific E2 node implementations used in our prototype are detailed in Sec.~\ref{sec:implementation}.

A typical pipeline execution proceeds through four main steps, as illustrated in Fig.~\ref{fig:manatee}. In \textbf{Step 1}, the \gls{smo} layer components are involved: the Template Server retrieves the pipeline template, the xApp Registry provides the xApp definition, and the Controller persists the pipeline configuration to the \gls{cicd} Registry. The Controller orchestrates the remaining steps by sending commands and retrieving information from the other layers. \textbf{Step 2} deploys or updates new xApp versions and undeploys old or failed revisions within the xApp layer. These operations are performed via the API Server and are stored in the Object Registry of the target platform (Kubernetes or OpenShift), while the Controller Manager ensures that resources remain in their desired state. In \textbf{Step 3.1}, the Controller configures the Service Mesh Control Plane through the Service Mesh API, storing routing rules and policies in the respective registries. These configurations are then propagated to the Service Mesh Data Plane in \textbf{Step 3.2}, where the sidecar proxies are dynamically configured. Here, the Observability Framework and the Controller work together to maintain the desired state of the mesh. Finally, in \textbf{Step 4}, once the xApps are correctly deployed and configured in the \nearrtric, they subscribe to indication messages from the Target E2 Nodes.

\section{MANATEE Lifecycle Management and Testing Flows}
\label{sec:testingflows}

\oursol provides novel lifecycle management and testing flows to support testing in production of xApps, enabling the (i) migration of xApp control over E2 nodes (Sec.~\ref{subsec:micgrationxappcontrol}), (ii) Canary Deployment of xApps (Sec.~\ref{subsec:canarydeployment}), and (iii) A/B Testing of xApps (Sec.~\ref{subsec:abtesting}).
Upon \oursol deployment, the Kubernetes/OpenShift cluster is configured, the service mesh is installed, and the \nearrtric and xApp namespaces are created and labeled for automatic sidecar injection. Then, the \nearrtric components are deployed.
Across all lifecycle management and testing flows, failures at any phase require reviewing and correcting the relevant configuration files (e.g., Dockerfiles, Helm charts, Kubernetes manifests). Currently, this troubleshooting process is manual; however, the observability stack (logs, metrics, traces) and service mesh dashboard assist operators in diagnosing issues efficiently.

\begin{figure}[!ht]
\centering
\includegraphics[width=\columnwidth]{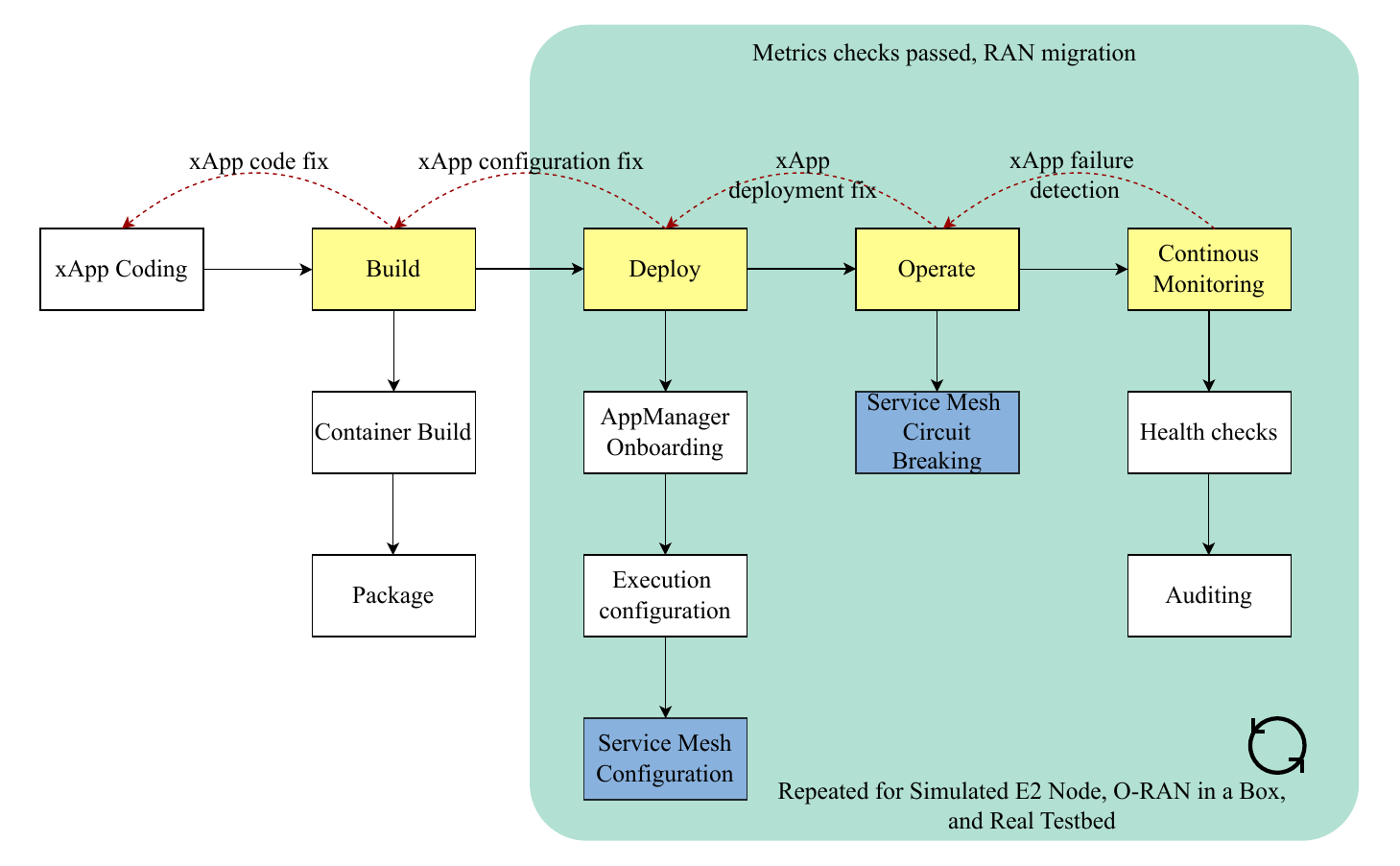}
\caption{Flow for migrating xApp control over E2 nodes.}
\label{fig:manatee-ran-migration}
\end{figure}

\subsubsection{Migration of xApp control over E2 nodes}\label{subsec:micgrationxappcontrol}

Figure~\ref{fig:manatee-ran-migration} shows the typical stages for the migration of the xApp from the testing environment to pre-production and production. 
This flow describes the progressive testing approach an xApp follows to ensure proper validation before deployment in a real \gls{ran}, thereby avoiding system disruptions and unexpected behaviors. The process begins by connecting the xApp to node, then migrating control to an emulated E2 node, and finally connecting to a production E2 node. The specific implementations of these E2 nodes (e.g., software simulators, containerized emulators, or physical base stations) are discussed in Sec.~\ref{sec:implementation}.
Once the xApp code is delivered, it enters the \emph{build phase}, which encompasses creating the container image and packaging it into a Helm chart. The Helm package is then distributed to a private repository for deployment.

The \emph{deploy phase} involves several steps. First, the xApp pod is created and started. The \gls{gnb} identifier is then passed to ensure the xApp connects to the correct E2 node. Next, xApp onboarding occurs, typically handled within the application itself, though it can also be performed separately via \gls{api} calls to the xApp Manager component within \nearrtric. Following this, appropriate VirtualServices and DestinationRules are configured to enable traffic routing to the xApp. 

Once the deployment succeeds, the xApp enters its \emph{operate phase}, where all components should function correctly and be visible in the service mesh dashboard. Common issues at this stage include the xApp not appearing in the xApp Manager's registered list, mesh connectivity problems, or improper xApp execution startup. Most of these issues can be resolved through runtime adjustments.

Finally, the \emph{continuous monitoring phase} validates that all health probes function correctly. xApp execution results are persisted to enable auditing, while Prometheus local storage captures logs, metrics, and traces for comprehensive observability.  The metrics collected by Prometheus include infrastructure-level telemetry (CPU, memory, network, pod health). E2 KPMs received by the xApp are processed and stored in InfluxDB for time-series analysis of RAN performance 
indicators. The upgrade of the xApp to a new E2 node passes through a manual gate to ensure the xApp behavior is properly caught and audited.

This lifecycle management and testing flow leverages several \oursol components: the \emph{DevOps API} to trigger and manage the migration pipeline, the \emph{Template Server} to retrieve parametrized workflow definitions with the target E2 node configurations, the \emph{xApp Registry} to fetch xApp deployment specifications, and the \emph{Controller} to reconcile the desired state and coordinate transitions between E2 nodes. The \emph{Service Mesh Data Plane} handles traffic routing through VirtualServices and DestinationRules, while the \emph{Observability Framework} collects metrics to validate xApp behavior at each migration stage.

\begin{figure}[!ht]
\centering
\includegraphics[width=\columnwidth]{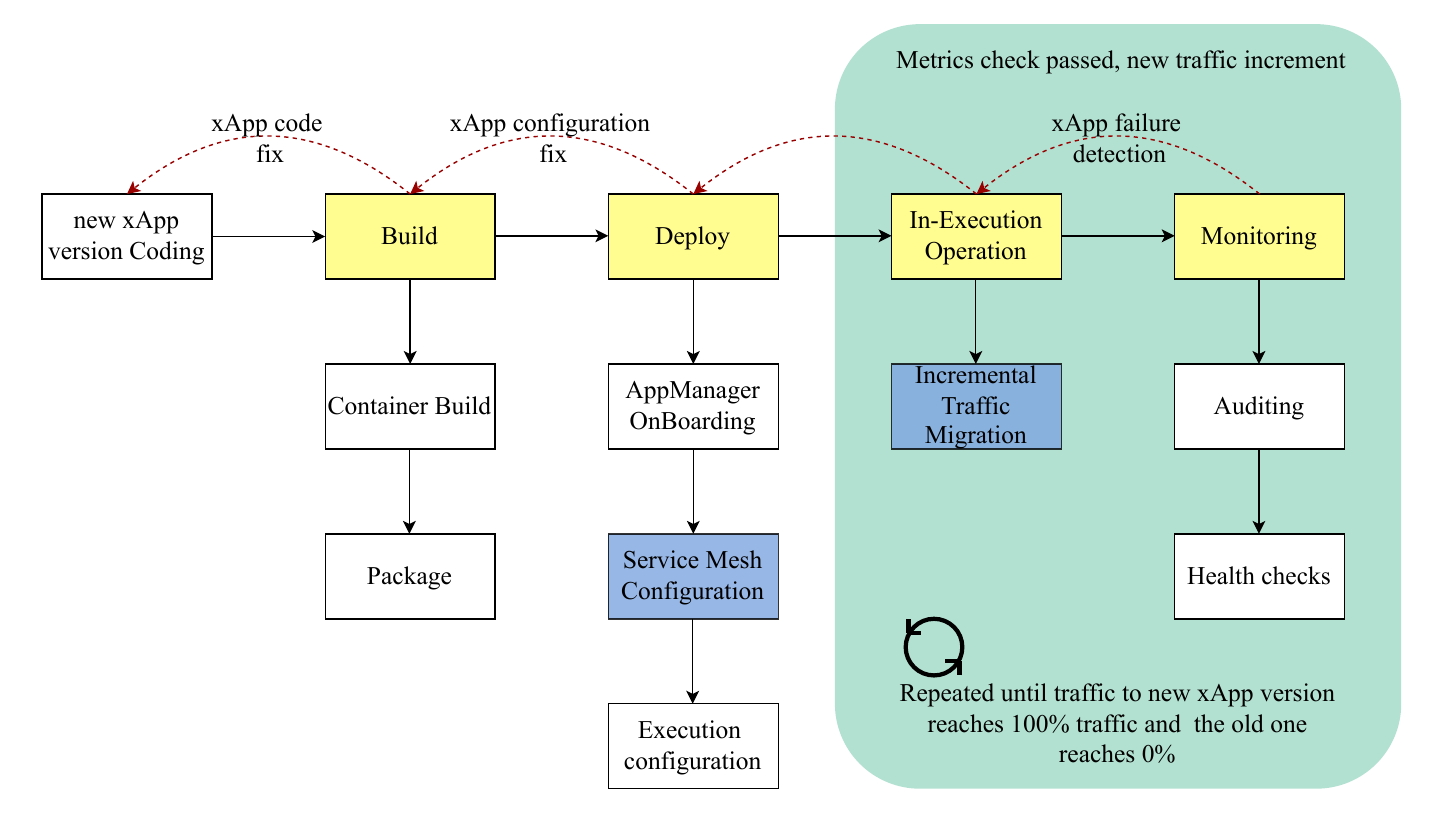}
\caption{Canary deployment of xApp to gradually shift traffic from the version in production to a new one.}
\label{fig:manatee-canary-deployment}
\end{figure}

\subsubsection{Canary Deployment of xApps}\label{subsec:canarydeployment}

Figure~\ref{fig:manatee-canary-deployment} shows the canary deployment flow in \oursol. This pipeline implements a canary deployment strategy to progressively introduce a new xApp version into the production environment while mitigating risks of unexpected behavior under varying traffic conditions. The process begins with a stable xApp version (e.g., v1) handling production traffic, then gradually shifts load to a new version (e.g., v2) through configurable traffic increments. This approach is particularly suited for deploying new xApp implementations or significant feature additions that require careful validation before full adoption in production. The traffic migration steps, defining the percentage increments and validation criteria, are configured before the pipeline execution. While this pipeline assumes connection to a real \gls{gnb}, the E2 node target remains configurable via the \gls{gnb} ID parameter, enabling flexibility in deployment scenarios. 

Once the new xApp code is delivered, it enters the \emph{build phase}, encompassing container image creation and Helm chart packaging. The Helm package is distributed to a private repository for deployment. 

The \emph{deploy phase} involves several sequential steps. The new xApp pod (v2) is created and started alongside the existing one (v1). The \gls{gnb} identifier is passed to ensure proper E2 node connectivity. xApp onboarding occurs either within the application or via \gls{api} calls to the xApp Manager component of the \nearrtric. Critical to canary deployment, VirtualServices and DestinationRules are configured to initially route 100\% of traffic to v1 and 0\% to v2. 

Upon successful deployment, the \emph{operate phase} begins with progressive traffic migration. Traffic to v2 increases incrementally according to the pre-configured steps (e.g., 10\%, 25\%, 50\%, 75\%, 100\%), while v1 traffic decreases accordingly. At each increment, the system validates v2 stability before proceeding. Issues in this phase are the same as those in the previous testing flow. 
For stateful xApps that maintain internal state (e.g., KPI trends, learned parameters, UE context), state migration or synchronization mechanisms may be needed to ensure v2 operates with consistent context. The traffic increment strategy should account for state warm-up periods to avoid initial performance degradation.

The \emph{continuous monitoring phase} validates health probes at each traffic increment. If v2 exhibits degraded performance, errors, or fails health checks, an automatic rollback to v1 preserves system stability, and the v2 deployment is aborted. Conversely, if v2 successfully handles all traffic increments, it gradually assumes full production load, and v1 is decommissioned. Throughout this process, execution results are persisted for auditing, while Prometheus captures logs, metrics, and traces, providing comprehensive observability for validation and troubleshooting.

This lifecycle management and testing flow involves multiple \oursol components: the \emph{DevOps API} exposes endpoints to configure and execute the canary pipeline, while the \emph{Template Server} provides parametrized workflow definitions specifying traffic increment percentages and validation thresholds. The \emph{\gls{cicd} Registry} stores the canary deployment pipeline definition, and the \emph{xApp Registry} maintains both xApp versions for deployment. The \emph{Controller} orchestrates progressive traffic shifts by updating VirtualServices and DestinationRules in the \emph{Service Mesh Data Plane}, and triggers automatic rollbacks when anomalies are detected. The \emph{Observability Framework} continuously feeds health metrics to the Controller for stability validation at each traffic increment.

\begin{figure}[!ht]
\centering
\includegraphics[width=\columnwidth]{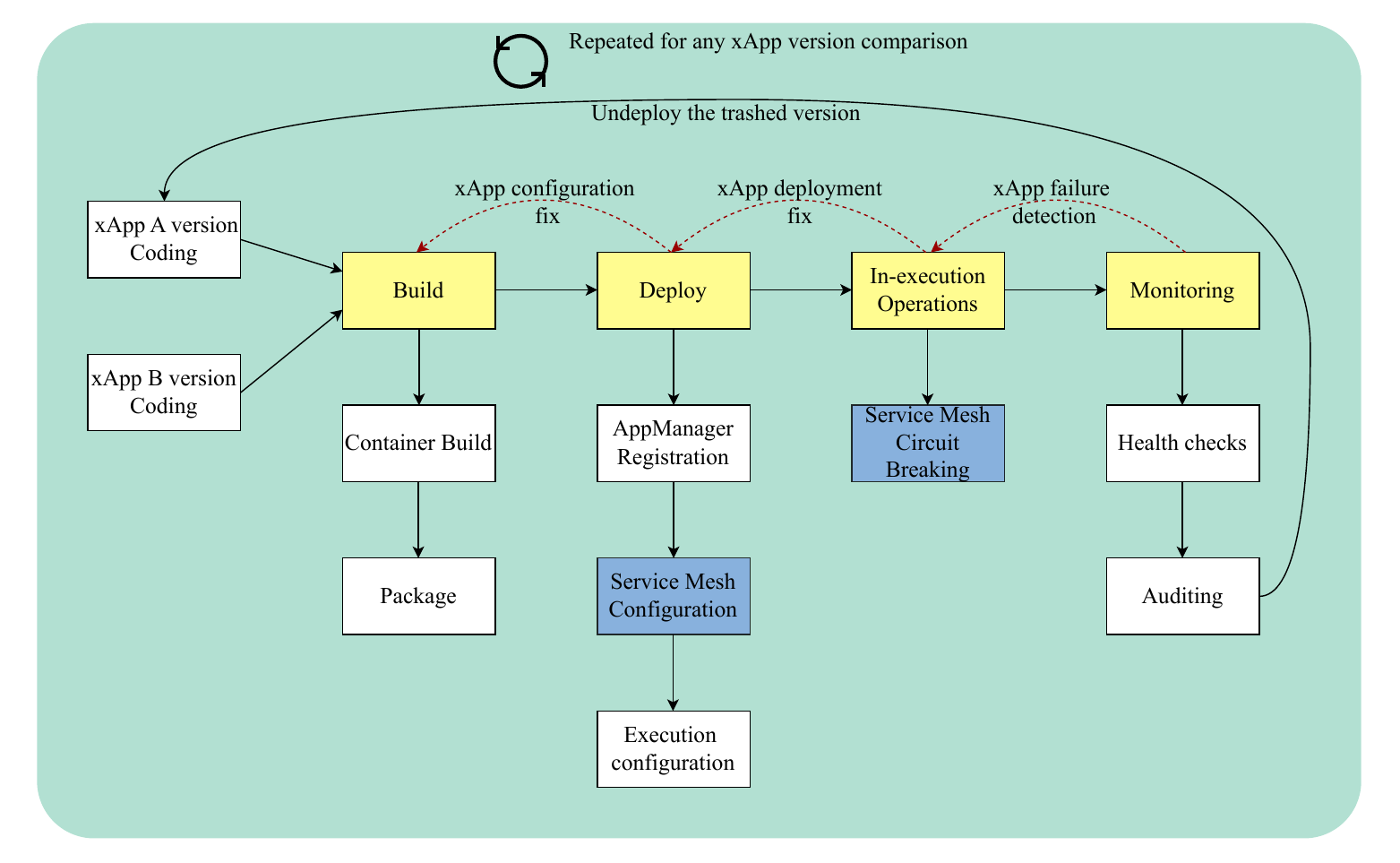}
\caption{A/B testing flow to enable the concurrent execution of different xApp versions.}
\label{fig:manatee-AB-testing}
\end{figure}

\subsubsection{A/B Testing of xApps}\label{subsec:abtesting}

Figure~\ref{fig:manatee-AB-testing} shows the execution flow of A/B testing in \oursol. This pipeline implements an A/B testing strategy to compare two xApp versions (A and B) that target the same control objective but employ different implementations or optimization strategies. The pipeline enables the concurrent operation of both versions under controlled conditions to evaluate performance, efficiency, and behavioral differences through statistical comparison of telemetry data. 
This approach is particularly suitable for evaluating competing optimization algorithms, resource allocation strategies, or control policies within xApps that have already undergone extensive validation. Proper A/B testing requires isolation between variants to ensure fair comparison. For this reason, either the xApps connect to equivalent E2 node instances with comparable network and traffic conditions, or they can operate on separate network slices (if slice-aware) with similar traffic profiles. Unlike canary deployments focused on risk mitigation, A/B testing prioritizes comparative performance analysis over progressive rollout. 

The process begins by deploying both xApp versions simultaneously. Once the xApp code for each variant is delivered, both enter the \emph{build phase}, encompassing container image creation and Helm chart packaging. Each Helm package is distributed to a private repository with distinct version tags. 

The \emph{deploy phase} provisions both xApp pods concurrently. Each receives its designated \gls{gnb} identifier or slice configuration to ensure proper E2 node connectivity and traffic isolation. xApp onboarding occurs either within the applications or via \gls{api} calls to the xApp Manager component of the \nearrtric. Critical to A/B testing, VirtualServices and DestinationRules are configured to route traffic according to the experimental design, typically based on slice identifiers, UE characteristics, or request attributes to ensure randomization and balance. 

Once the deployment succeeds, both xApps enter the \emph{operate phase} and execute their control and/or monitoring tasks concurrently. All components are be visible in the service mesh dashboard, enabling real-time comparison of traffic patterns, latency, and resource utilization. If anomalous behavior emerges in either variant, circuit-breaking mechanisms in DestinationRules can isolate the problematic xApp by preventing its messages from reaching the E2 termination, protecting network stability while allowing the experiment to continue with the functioning variant. Common issues include xApps missing from the xApp Manager's registered list, mesh connectivity failures, or startup errors, and are resolvable through runtime adjustments.

The \emph{continuous monitoring phase} validates health probes for both versions while collecting comprehensive telemetry. Prometheus captures logs, metrics, and traces from both xApps, enabling statistical analysis of \glspl{kpi} such as throughput, latency, resource efficiency, and control effectiveness. 
Upon experiment completion, the saved results are used for statistical significance testing to determine which xApp variant is performing better. Then, this version is retained and scaled to handle all traffic (across both slices or E2 nodes), while the worst-performing version is decommissioned along with its associated infrastructure. This data-driven approach ensures optimization decisions are grounded in empirical evidence.

This lifecycle management and testing flow utilizes several \oursol components: the \emph{DevOps API} enables configuration of the A/B experiment parameters, including test duration and success criteria. The \emph{Template Server} stores workflow definitions with placeholders for slice identifiers and \gls{kpi} thresholds, while the \emph{\gls{cicd} Registry} maintains the A/B testing pipeline definition. The \emph{xApp Registry} hosts both xApp variants with distinct version tags. The \emph{Service Mesh Data Plane} enforces traffic isolation between variants through VirtualServices and DestinationRules, and enables circuit-breaking for fault isolation. The \emph{Controller} monitors experiment progress and, upon completion, coordinates the decommissioning of the underperforming variant. The \emph{Observability Framework} is central to this flow, collecting telemetry from both variants to enable statistical comparison of \glspl{kpi}.

%%%%%%%%%%%%%%%%%%%%%%%%%%%%%%%%%%%%%%%%%%%%%%%%%%%%%%%%%%%%%%%%%%%%%%%%%%%%%%%%%%%%%%%%%%%%%%%%%%%%%%%%%%%%%%%%%%
%%%%%%%%%%%%%%%%%%%%%%%%%%%%%%%%%%%%%%%%%%%%%%%%%%%%%%%%%%%%%%%%%%%%%%%%%%%%%%%%%%%%%%%%%%%%%%%%%%%%%%%%%%%%%%%%%%
%IMPLEMENTATION
%%%%%%%%%%%%%%%%%%%%%%%%%%%%%%%%%%%%%%%%%%%%%%%%%%%%%%%%%%%%%%%%%%%%%%%%%%%%%%%%%%%%%%%%%%%%%%%%%%%%%%%%%%%%%%%%%%
%%%%%%%%%%%%%%%%%%%%%%%%%%%%%%%%%%%%%%%%%%%%%%%%%%%%%%%%%%%%%%%%%%%%%%%%%%%%%%%%%%%%%%%%%%%%%%%%%%%%%%%%%%%%%%%%%%

\section{MANATEE Prototype}\label{sec:implementation}

\begin{figure*}[!ht]
\centering
\includegraphics[width=\textwidth]{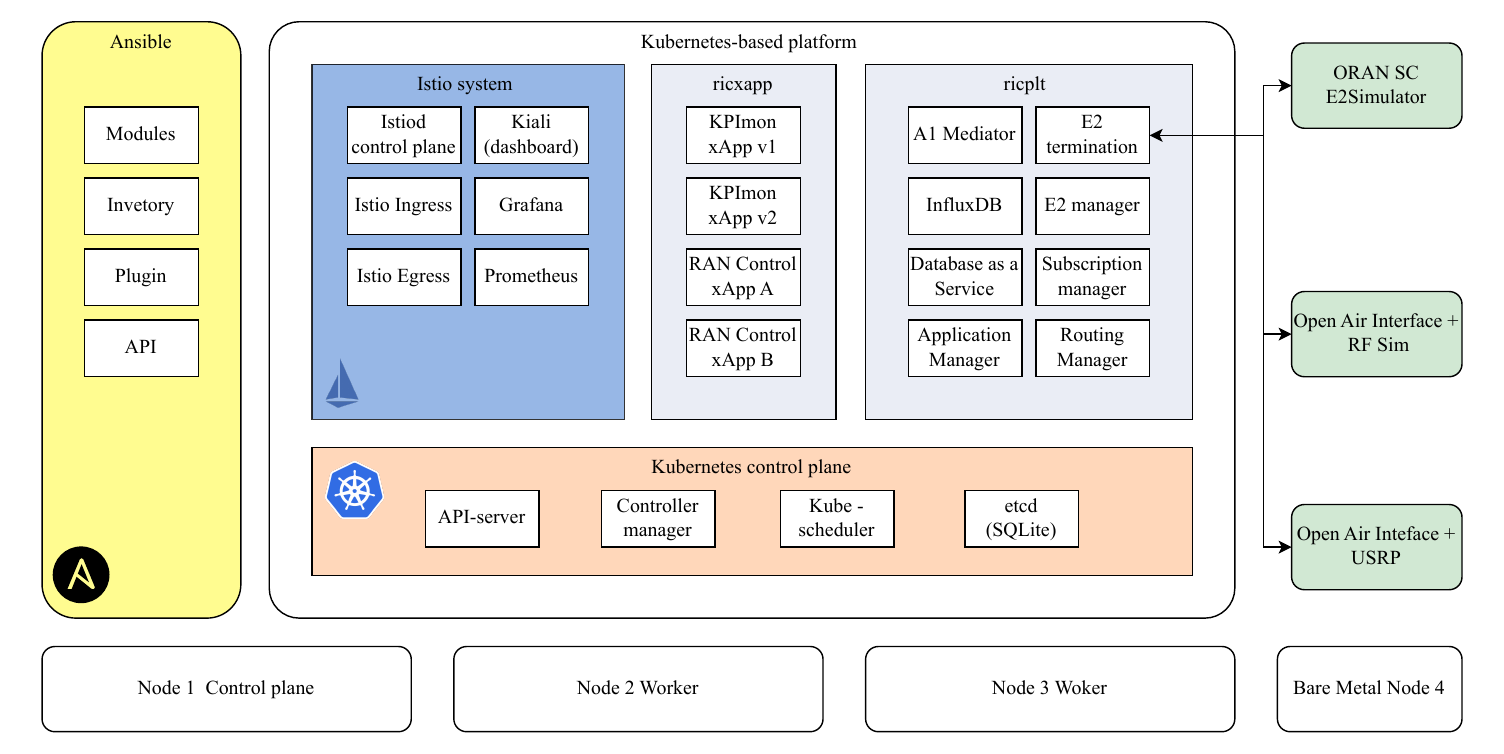}
\caption{Logical diagram of \acrshort{oursol} prototype.}
\label{fig:manatee-implementation}
\end{figure*}

We prototype \oursol following the architecture illustrated in Fig.~\ref{fig:manatee-implementation}. We instantiate the \gls{oransc} \nearrtric on a three-node Kubernetes K3s cluster~\cite{k3s}. Each node has an 8-core CPU and 32 GB of RAM, running Linux kernel 5.19.0-41-generic and Ubuntu Server 22.04. We leverage Helm for package management of \gls{ric} components (version: 3.19), Istio (demonstration profile) for service mesh capabilities (version: 1.28.0), and Ansible for \gls{cicd} orchestration through declarative playbooks (version: 2.18.0). The Ansible control plane operates co-located with the Kubernetes control-plane node. With respect to reference xApp implementations, we rely on a \gls{rc} xApp that can control the amount of \glspl{prb} allocated to the \gls{ran}~\footnote{https://hub.docker.com/r/angeloferaudo/prb-control-xapp} and to the KPI monitoring one~\footnote{https://github.com/o-ran-sc/ric-app-kpimon-go.git}.
During the mesh comparison tests, we also deployed Istio Ambient mode version 1.28.0 and Cilium 1.18.0.
Then, we integrate \oursol with different E2 node implementations: (i)~the \gls{oransc} E2 simulator for synthetic traffic generation; (ii)~the ORANInABox solution\footnote{https://github.com/aferaudo/ORANInABox.git} comprising \gls{oai} \gls{gnb} and \gls{ue} components; and (iii)~an \gls{oai} deployment interfaced with an NI USRP X310 \gls{sdr} for real-world validation. In the following, we provide details regarding each lifecycle management and testing flow implementation.

The prototyping effort revealed several technical challenges inherent to the O-RAN SC platform~\cite{oransc}. First, the \gls{rmr} implementation proved fragile, with routing table generation frequently causing xApp communication failures due to inconsistent table propagation and limited debugging capabilities. Second, integrating service mesh technologies with \gls{rmr} required careful configuration, as \gls{rmr} is a custom messaging protocol that does not conform to HTTP/gRPC patterns; consequently, we applied the mesh at Layer4, optimizing it for TCP traffic rather than relying on the Layer7 capabilities that Istio and Cilium typically leverage. This necessitated traffic policy exceptions and motivated our evaluation of multiple mesh implementations, as well as a different A/B testing strategy. Third, the xApp ecosystem remains immature: few reference implementations provide functional E2 control capabilities, mainly those derived from xDevSM~\cite{xdevsmferaudo2024}, and those that exist are tightly coupled to specific Helm chart structures, limiting deployment flexibility.

\textbf{Migration of xApp Control over E2 Nodes.}
The Ansible orchestration framework implements a three-phase deployment pipeline spanning simulated, emulated (ORANInABox), and real testbed environments. Global variables specify Kubernetes namespaces, Helm repositories, PLMN identifiers, timeout thresholds, and xApp parameters. Each phase executes: PLMN discovery and validation via E2 Manager APIs, xApp instantiation through Helm with phase-specific releases, pod identification, process initialization, Application Manager registration, health verification, and manual approval gates. For the simulated and emulated phases, asynchronous approval mechanisms trigger controlled tear-down procedures, enabling xApp redeployment for the following testing stages. All phases incorporate exception handling with automatic Helm cleanup upon failure to ensure a consistent system state.

\textbf{Canary Deployment of xApps.}
The Canary Deployment pipeline implements a phased release strategy for xApp version transition through optional image build, parallel deployment of baseline and candidate versions, and Istio service mesh configuration via Virtual Services and Destination Rules. The canary rollout performs gradual traffic shifting, incrementally transferring weight to the candidate version. Each reallocation step triggers validation of health probes and telemetry metrics (e.g., latency), enabling stage-gated progression where performance assessment at each traffic percentage step authorizes advancement to subsequent phases.

\textbf{A/B Testing for xApps.}
The A/B testing pipeline implements egress-controlled testing through specialized roles for xApp provisioning, egress policy configuration, traffic generator deployment, and result collection. Pre-execution tasks verify the Kubernetes and Istio prerequisites, while the main workflow comprises xApp A/B deployment, application of egress policies, and temporal test execution, with results organized for comparative analysis.
The implementation addresses control serialization over identical \gls{ran} infrastructure through three modes: (i)~temporal multiplexing via service mesh circuit breaking for time-sliced alternation; (ii)~parallel slice-based operation with isolated control paths to distinct slices on the same \gls{gnb}; and (iii)~simultaneous multi-\gls{gnb} scenarios with independent associations. \oursol adopts the first mode as it is the most general approach and does not assume the availability of a large number of \glspl{gnb} or the use of slicing technologies.

%%%%%%%%%%%%%%%%%%%%%%%%%%%%%%%%%%%%%%%%%%%%%%%%%%%%%%%%%%%%%%%%%%%%%%%%%%%%%%%%%%%%%%%%%%%%%%%%%%%%%%%%%%%%%%%%%%
%%%%%%%%%%%%%%%%%%%%%%%%%%%%%%%%%%%%%%%%%%%%%%%%%%%%%%%%%%%%%%%%%%%%%%%%%%%%%%%%%%%%%%%%%%%%%%%%%%%%%%%%%%%%%%%%%%
%EXPERIMENTAL RESULTS
%%%%%%%%%%%%%%%%%%%%%%%%%%%%%%%%%%%%%%%%%%%%%%%%%%%%%%%%%%%%%%%%%%%%%%%%%%%%%%%%%%%%%%%%%%%%%%%%%%%%%%%%%%%%%%%%%%
%%%%%%%%%%%%%%%%%%%%%%%%%%%%%%%%%%%%%%%%%%%%%%%%%%%%%%%%%%%%%%%%%%%%%%%%%%%%%%%%%%%%%%%%%%%%%%%%%%%%%%%%%%%%%%%%%%

\section{Experimental Results}\label{sec:experimentalresults}

In this section, we evaluate the capabilities of \oursol across the lifecycle management and testing flows presented in Sec.~\ref{sec:testingflows}, as well as system metrics for different service mesh configurations. Section~\ref{subsec:servicemeshes} compares the latency introduced by various mesh architectures under burst and incremental traffic profiles, while Sec.~\ref{subsec:servicemeshoverhead} examines the overhead of Kubernetes \gls{crud} operations across mesh configurations. Section~\ref{subsec:testmigration} evaluates the latency of each phase in the xApp migration pipeline across simulated, emulated, and production environments. Section~\ref{subsec:testcanary} presents canary deployment tests, analyzing throughput variance over time along with latency and error rates under varying load conditions. Finally, Sec.~\ref{subsec:testabtesting} demonstrates A/B testing capabilities, showing ingress and egress throughput patterns using circuit-breaking mechanisms.

\subsection{Service Mesh Performance Comparison}\label{subsec:servicemeshes}

\begin{figure}[!ht]
\centering
\includegraphics[width=\columnwidth]{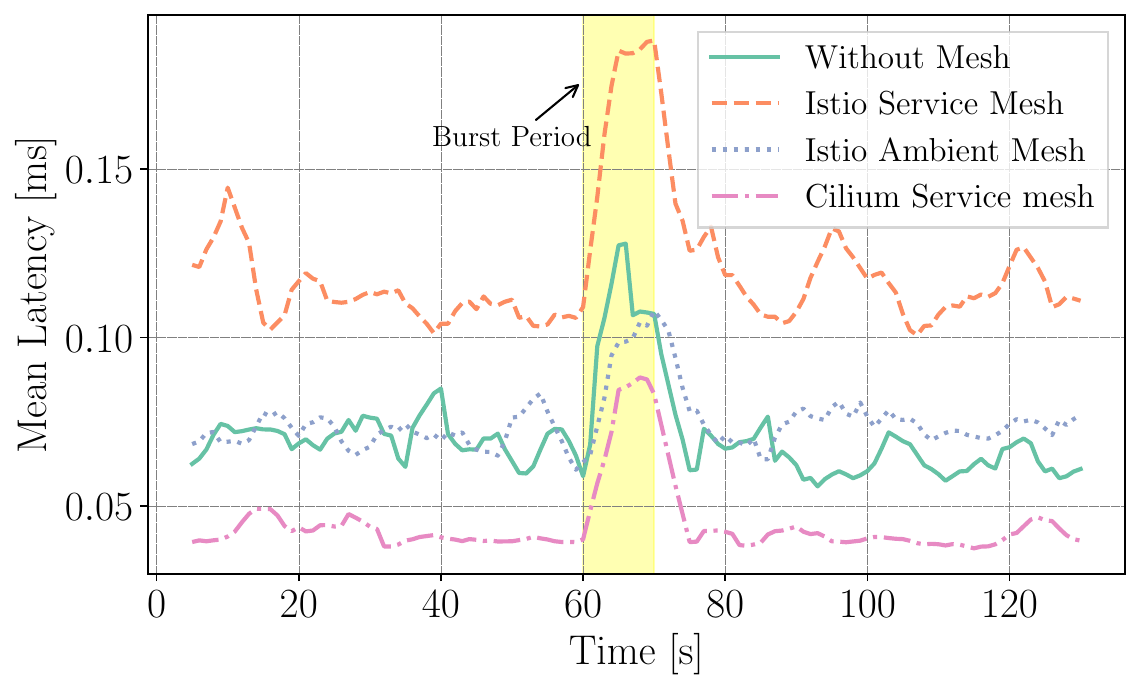}
\caption{Burst traffic test assessing performance of service meshes with different architectures.}
\label{fig:latency-burst-manatee}
\end{figure}

\begin{figure}[!ht]
\centering
\includegraphics[width=\columnwidth]{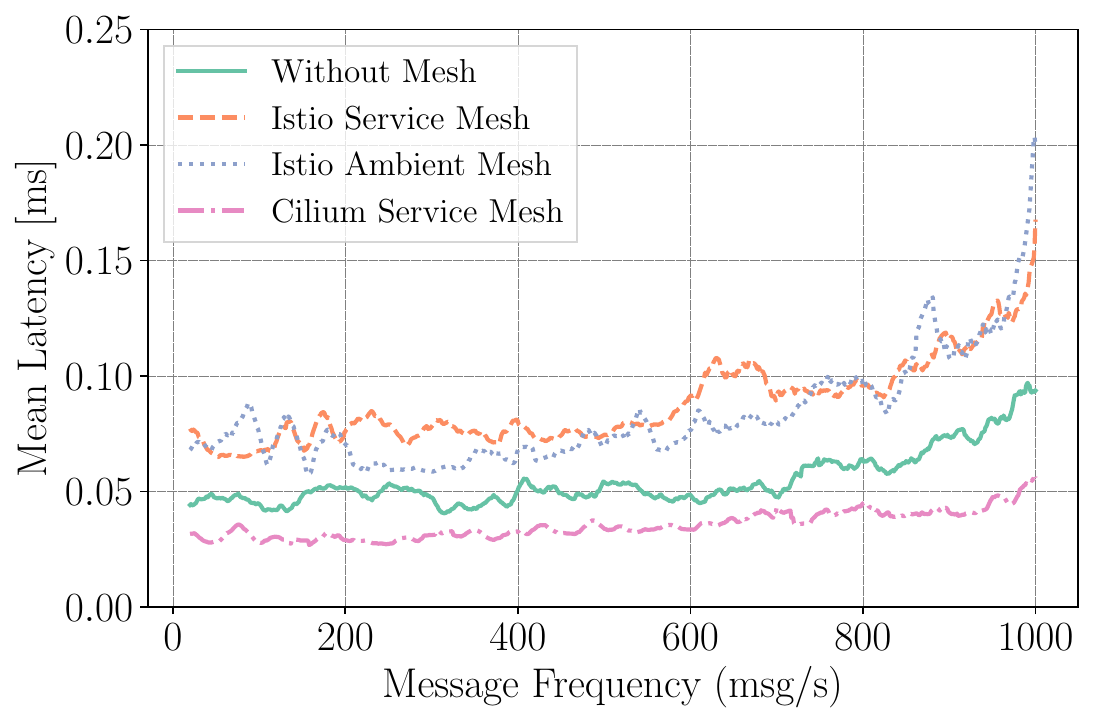}
\caption{Incremental traffic test assessing performance of service meshes with different architectures.}
\label{fig:latency-incremental-manatee}
\end{figure}

As mentioned in Sec.~\ref{sec:implementation}, there are several open-source service mesh projects supported by the \gls{cncf}~\cite{cncf}. In this evaluation, we compare the most relevant solutions to justify our implementation choice for \oursol. We consider four different approaches: (i)~a sidecar proxy-based model using Istio service mesh~\cite{istio}; (ii)~a sidecarless model with Istio Ambient Mesh~\cite{ambient}; (iii)~a sidecarless, eBPF-based in-kernel processing model using Cilium~\cite{cilium}; and (iv)~a no-mesh approach.

We test the mesh networking capabilities under different traffic profiles. Since most interactions within the \nearrtric occur through \gls{rmr}---a publish-subscribe protocol based on TCP---we compare the TCP performance of xApps under burst and incremental traffic profiles.
First, we stress-test our xApp with a burst traffic profile: requests flow at $100$\:msg/s for $60$\:s, then switch to $1000$\:msg/s for $20$\:s, and finally return to $100$\:msg/s.
Figure~\ref{fig:latency-burst-manatee} shows the latency over time for our four deployment configurations, showing an increase in latency upon the increase in request rate to $1000$\:msg/s after $60$\:s.
Results indicate that Istio service mesh is $61$\% slower than Istio Ambient mode and the no-mesh deployment. In contrast, Ambient mode exhibits only an $11$\% average performance degradation compared to the no-mesh baseline. In comparison, no mesh is $60$\% slower than Cilium.

We evaluate the mesh under an incremental traffic profile in which requests start at $1$\:msg/s and increase to $1000$\:msg/s, with an increment of $1$ message every second.
Figure~\ref{fig:latency-incremental-manatee} shows the latency as a function of the message frequency, which also corresponds to the elapsed time in seconds from the beginning of the experiment.
In this case, Istio service mesh is 11\% slower than Ambient Mesh, which is 53\% slower than the no-mesh approach. At the same time, the no-mesh approach is 56.12\% slower than Cilium.
In general, across both tests, the Istio service mesh introduces an overhead of between $60$-$75$\% compared to the test without a mesh. However, the additional latency remains well below $1$\:ms and does not significantly impact performance for the considered traffic types.
Finally, Cilium behaves the best, even better than the no-mesh approach, by performing in-kernel packet processing, reducing context switches, and prioritizing networking operations within the pod. 

\begin{comment}
Bursty Average Latency Difference Percentages:
No Mesh - Cilium: 60.49\%
Ambient - No Mesh: 10.80\%
Istio - Ambient: 60.98\%

Incremental Average Latency Difference Percentages:
No Mesh - Cilium: 56.12\%
Ambient - No Mesh: 52.52\%
Istio - Ambient: 10.75\%
\end{comment}

\subsection{Pipeline-Service Mesh Interaction}\label{subsec:servicemeshoverhead}

\begin{figure}[!ht]
\centering
\includegraphics[width=\columnwidth]{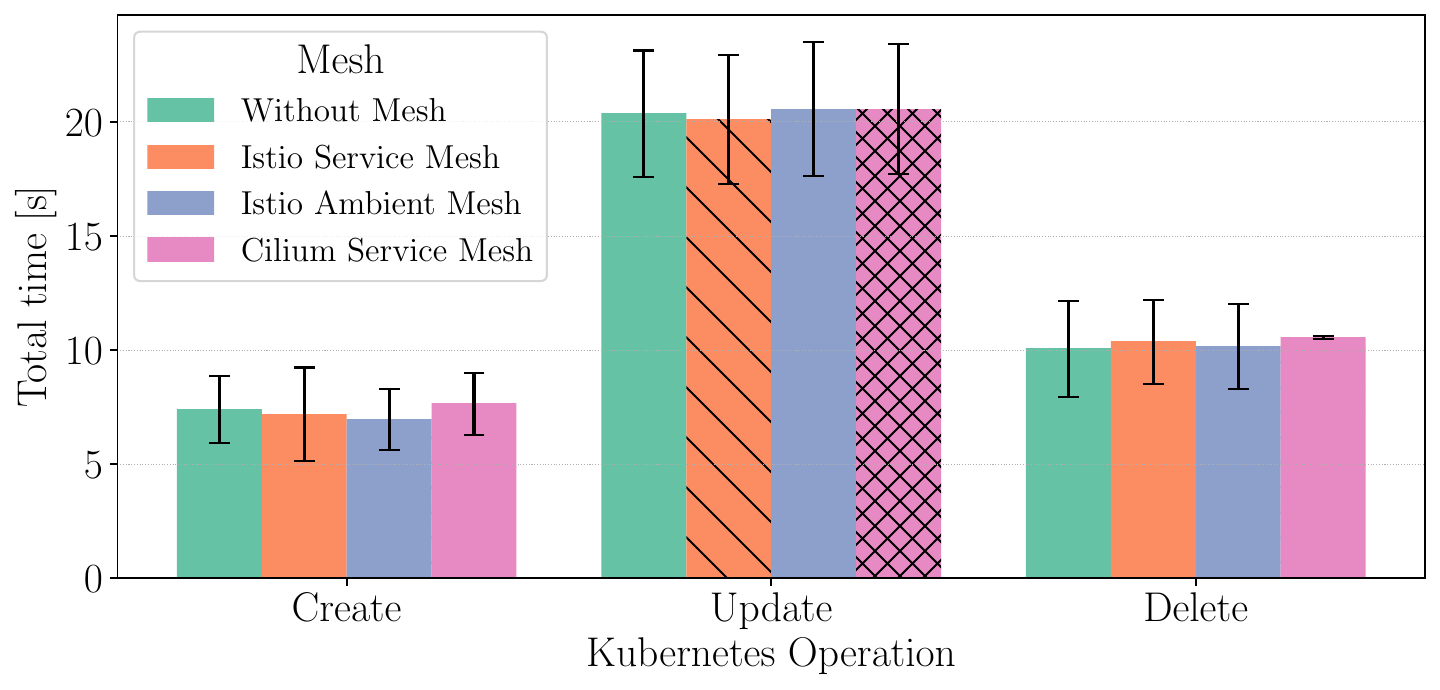}
\caption{Create, Update, Delete deployment operations comparison in Kubernetes for the different service mesh technologies versus the no-mesh baseline.}
\label{fig:crud-mesh}
\end{figure}

We now evaluate the overhead introduced by service meshes when performing Create, Update, and Delete operations on xApp deployments. Specifically, we measure the time required to: (i)~create a new xApp instance; (ii)~update its YAML manifest; and (iii)~delete the xApp pod. Each experiment run is repeated 100 times to ensure statistical significance.
From Fig.~\ref{fig:crud-mesh}, we notice that differences between mesh and no-mesh configurations are marginal across all three operations. Create operations range from $6.9$\:s (Istio Ambient) to $7.6$\:s (Cilium), representing a maximum overhead of around $10$\% compared to the no-mesh baseline ($7.3$\:s). Update operations exhibit consistent performance ($20.0$ to $20.5$\:s), which confirms that configuration updates, primarily involving API server interactions and ConfigMap propagation, are independent of mesh presence. Delete operations show minimal impact ($10.1$ to $10.6$\:s), as Kubernetes performs parallel container termination and the dominant latency source is the enforcement of the container grace period rather than per-container teardown.

\subsection{Migration of xApp Control Over E2 Nodes}\label{subsec:testmigration}

\begin{figure}
\centering
\includegraphics[width=\columnwidth]{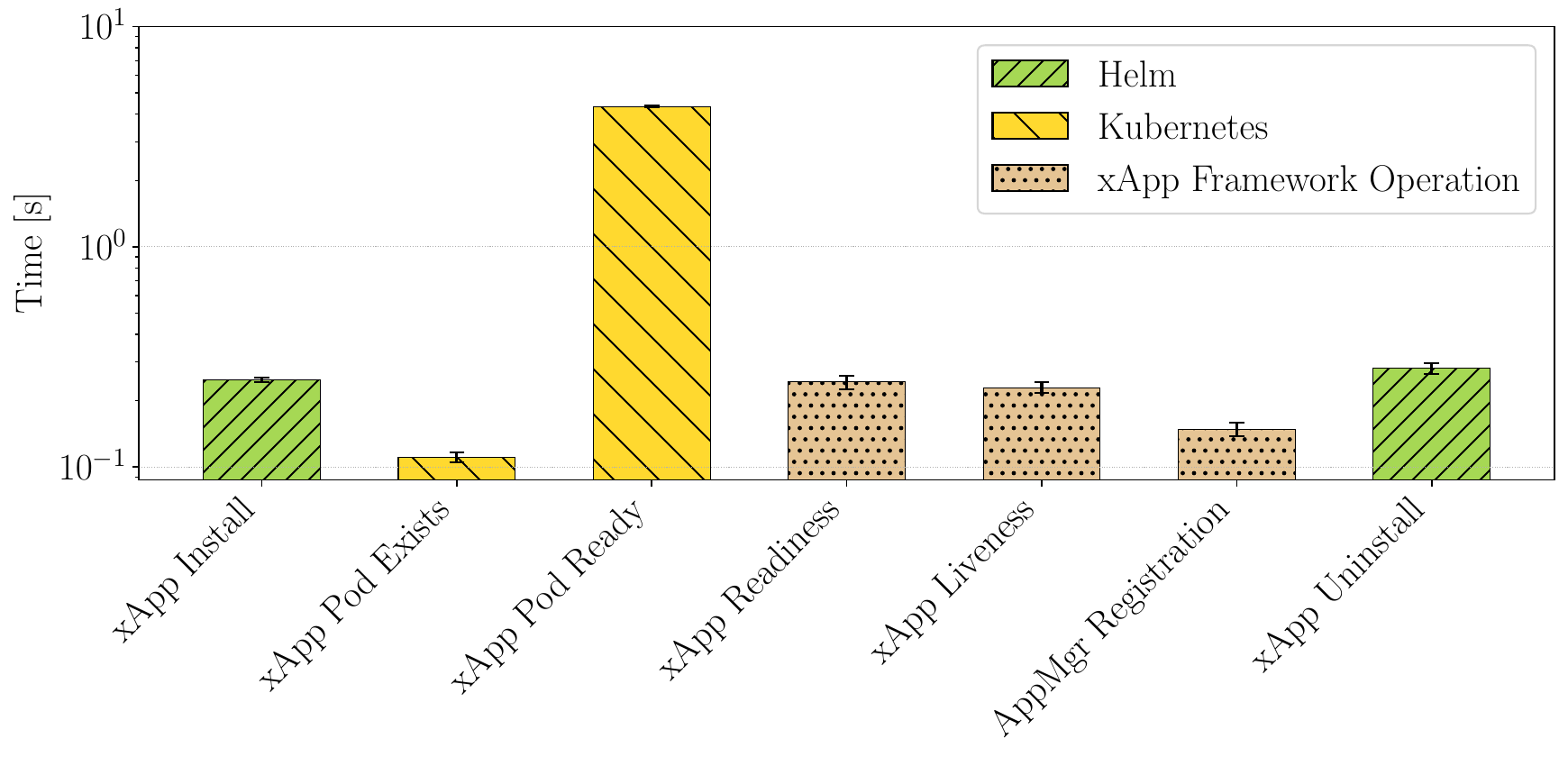}
\caption{Average execution time for each phase of the xApp migration pipeline, progressing from simulated E2 node control, to emulated gNB (ORANInABox), to production deployment.}
\label{fig:pipeline-benchmark}
\end{figure}

We now evaluate the capabilities of \oursol to progressively migrate xApp control across heterogeneous environments, enabling thorough validation before applying control to production \gls{ran} infrastructure that could potentially disrupt \gls{ue} connections or exhibit unintended behaviors. A representative use case involves an operator observing degraded performance in a deployed KPI-monitoring xApp and initiating a controlled migration to a previously validated, stable version; the pipeline ensures that the replacement xApp is revalidated across simulated and emulated environments before reaching production. The experiment in Fig.~\ref{fig:pipeline-benchmark} measures the time to complete this migration loop, assessing each deployment phase. The pipeline was implemented using Ansible and executed $100$ times to ensure statistical significance.

The pipeline executes in approximately $5.6$\:s per target gNB on average, introducing minimal overhead for validating xApp behavior across each environment. This metric represents the automated execution time for deploying, configuring, and validating the xApp within a single environment. This measurement excludes the duration of manual approval gates, which are incorporated between phases and require human validation before proceeding to the next E2 node. This human-in-the-loop latency is intentionally omitted from the reported metrics as it varies significantly between executions and operators.  In a complete migration scenario, the xApp goes through the pipeline three times, once for each target E2 node (simulated, emulated, and real), resulting in a total automated execution time of approximately $16.8$\:s.
The target endpoint selection is configurable via command-line parameters at xApp startup, enabling flexible deployment orchestration across diverse network topologies.

\subsection{Canary Deployment of xApps}\label{subsec:testcanary}

\begin{figure*}[!t]
\centering
\begin{subfigure}[b]{0.32\textwidth}
    \centering
    \includegraphics[width=\textwidth]{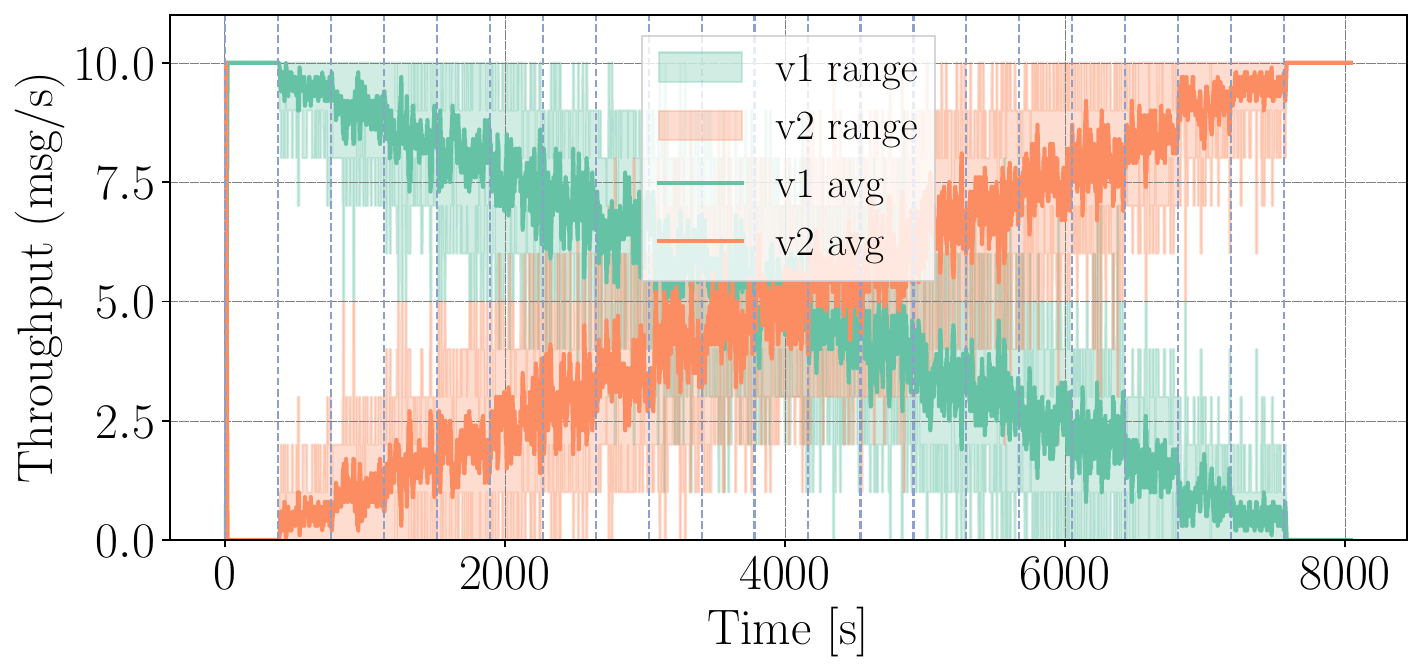}
    \caption{$10$\:msg/s load}
    \label{fig:canary10}
\end{subfigure}
\hfill
\begin{subfigure}[b]{0.32\textwidth}
    \centering
    \includegraphics[width=\textwidth]{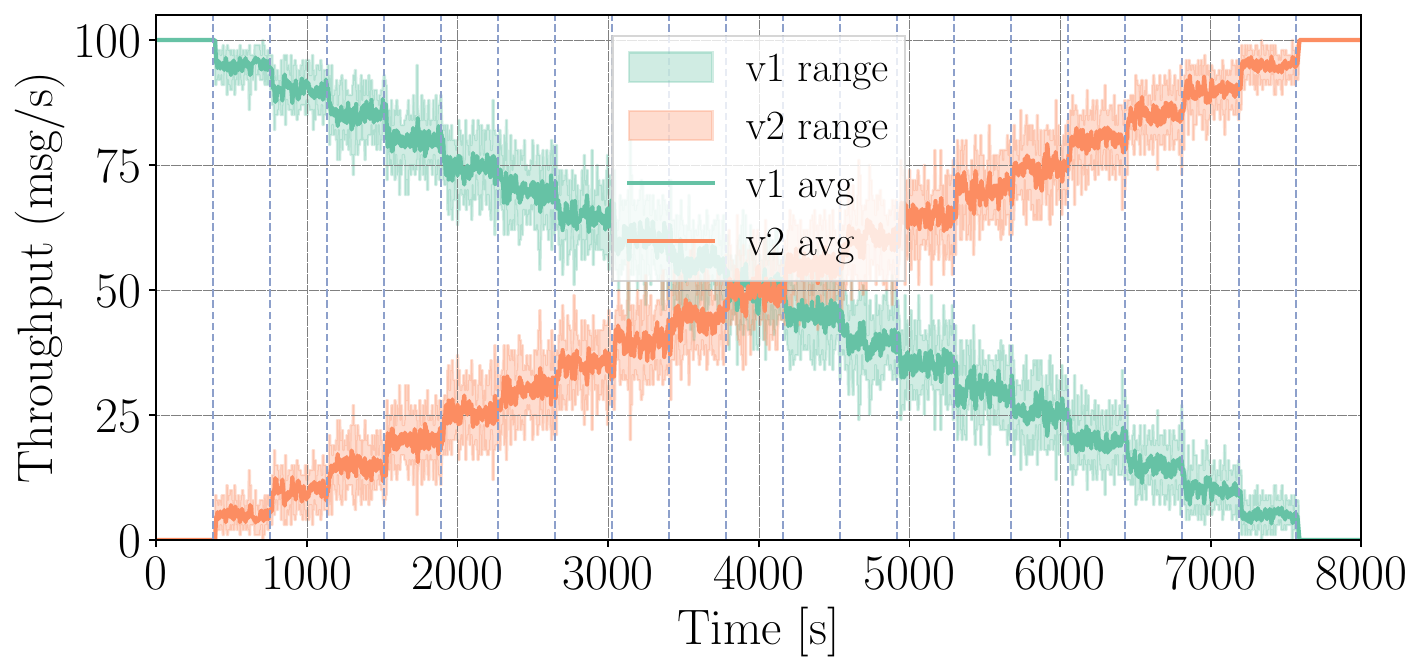}
    \caption{$100$\:msg/s load}
    \label{fig:canary100}
\end{subfigure}
\hfill
\begin{subfigure}[b]{0.32\textwidth}
    \centering
    \includegraphics[width=\textwidth]{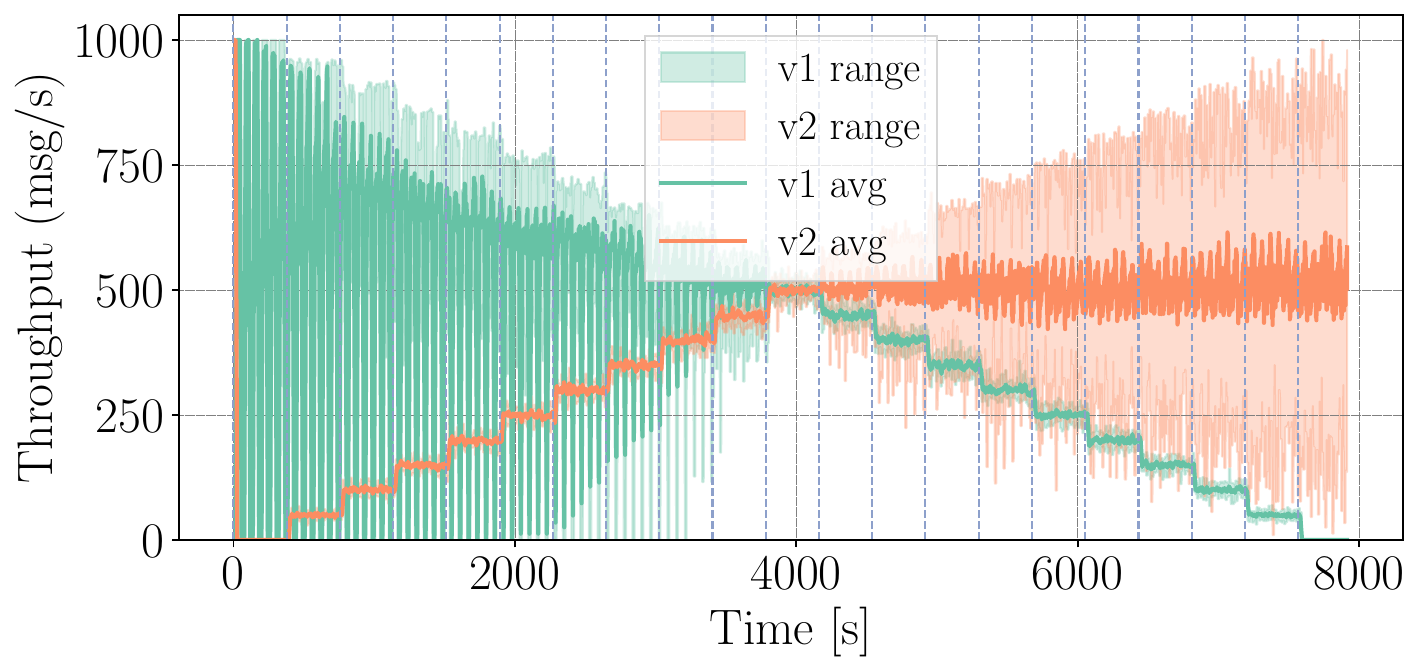}
    \caption{$1000$\:msg/s load}
    \label{fig:canary1000}
\end{subfigure}
\caption{Canary deployment test: $120$ minutes for migration from v1 to v2, with $5$\% step in migration every 6 minutes, at different message rates reaching the xApp.}
\label{fig:canary-throughput}
\end{figure*}

\begin{figure}[!ht]
\centering
\includegraphics[width=\columnwidth]{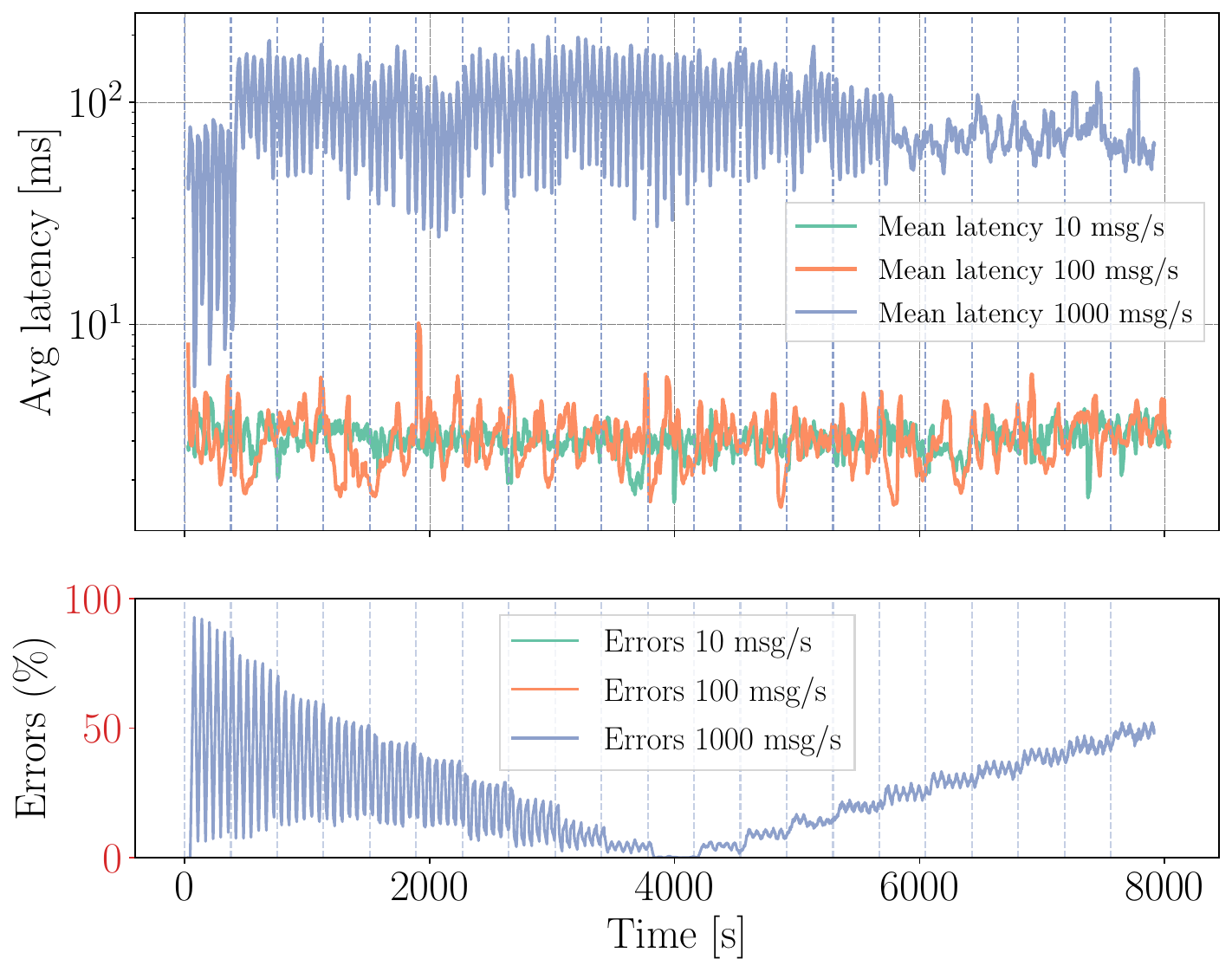}
\caption{Canary deployment test latency and error comparison for low ($10$\:msg/s), medium ($100$\:msg/s), and high load ($1000$\:msg/s) scenarios.}
\label{fig:canary-latency-error}
\end{figure}

In this set of experiments, we demonstrate the capabilities of \oursol to perform gradual canary deployments of xApps while maintaining service continuity under varying load conditions. We deploy two versions of the \gls{kpi} monitoring xApp from~\cite{xdevsmferaudo2024}---referred to as v1 and v2---which continuously monitor RAN performance metrics through E2 interface 
subscriptions. The xApp receives periodic E2 indication messages containing cell-level and UE-level measurements (e.g., throughput, resource utilization, packet loss rates) from base stations, processes these metrics, and stores them in an InfluxDB time-series database for real-time monitoring and historical analysis.

We use our Ansible playbook implementation integrated with Istio service mesh to orchestrate the deployment. To ensure statistical reliability, experiments for each test configuration are repeated 50 times. A mock E2 simulator is configured to generate indication messages at controlled rates to simulate realistic monitoring scenarios, allowing us to evaluate the xApp's ability to sustain different message processing loads during the canary deployment transition.

The canary deployment strategy we follow incrementally shifts traffic from v1 to v2 by increasing the percentage of requests directed to the v2 xApp by 5\% every 6 minutes, thus completing the full migration in approximately 2 hours. This gradual approach allows for early detection of potential issues with the new v2 version while minimizing degradation to the overall system. We evaluate the system behavior under three distinct load scenarios: low load ($10$\:msg/s), medium load ($100$\:msg/s), and high load ($1000$\:msg/s), representing realistic operational conditions for network monitoring applications \cite{RODRIGUES2025112500}. Results are shown in Fig.~\ref{fig:canary-throughput}.

Figure~\ref{fig:canary10} illustrates the low-load traffic migration pattern at $10$\:msg/s, where both xApp versions maintain a stable throughput transition with no noticeable degradations,demonstrating that \oursol can handle canary deployments effectively for this load level.
Figure~\ref{fig:canary100} presents results for the medium-load scenario at $100$\:msg/s. Similar to the previous case, the migration proceeds smoothly with both versions properly handling their allocated traffic. The throughput curves clearly distinguish between the v1 and v2 traffic distributions, validating the precision of the traffic-splitting mechanism provided by the Istio service mesh.
Finally, Fig.~\ref{fig:canary1000} depicts the high-load test at $1000$\:msg/s, representing a stress scenario for both xApp versions. We observe that, at this rate, the xApp cannot maintain the expected throughput for v1, whereas the throughput for v2 never reaches the target rate of $1000$\:msg/s. It is worth noticing that in scenarios where the xApp struggles to maintain throughput, the pipeline should immediately interrupt the deployment.
However, we deactivated this canary deployment stopping mechanisms to demonstrate the extent to which the system can be stressed in extreme cases. This $600$\:msg/s threshold corresponds to the saturation point of the  TCP connection, beyond which the combination of increased round-trip latency, socket buffer exhaustion, and message processing overhead leads 
to packet drops and retransmissions, manifesting as degradations in the deployment.

Figure~\ref{fig:canary-latency-error} provides a comparative analysis of latency and error rates across the low-, medium-, and high-load test scenarios. Latency measurements for the low-load ($10$\:msg/s) and medium-load ($100$\:msg/s) cases reveal that the mean latency remains relatively stable throughout the experiment, whereas in the high-load ($1000$\:msg/s) case it increases to approximately $100$\:ms. Notably, the error rate remains at or near $0$\% for both the low- and medium-load scenarios throughout the entire migration period. More errors, however, appear in the high-load scenario due to resource saturation, which typically triggers an automated rollback in the deployment pipeline. In general, the xApp demonstrates the ability to undergo a traffic load of around $600$\:msg/s, which also corresponds to $0$ errors in the error plot, typical for standard TCP connections. 

Overall, the above results demonstrate that \oursol enables reliable, automated canary deployments of xApps with fine-grained traffic control and minimal service disruption. The integration with Istio service mesh provides robust traffic management capabilities, while the Ansible-based automation ensures repeatability and consistency across deployments. The system ability to maintain performance stability during gradual version transitions under appropriate load conditions validates its suitability for production Open \gls{ran} environments where service continuity is critical.

\subsection{A/B Testing for xApps}\label{subsec:testabtesting}

This experiment demonstrates the capability of \oursol to perform A/B testing on two concurrent xApp versions, while preventing control conflicts through circuit-breaking mechanisms. We deployed two PRB control xApp, namely versions A and B, which periodically send E2 control messages to adjust radio resource allocation ratios for UEs at the base station. Both versions implement different allocation strategies and were evaluated under controlled traffic conditions to assess their resource management approaches.

Because the \gls{ric} uses a publish-subscribe architecture with no message retention, our A/B testing approach requires careful traffic orchestration. In this PRB control scenario, both xApp versions execute their allocation logic independently. However, egress traffic, representing E2 control messages sent to adjust PRB quotas at the \gls{ran}, was selectively enabled for only one xApp version at a time using the Istio circuit-breaking capabilities. This prevents conflicts that would otherwise arise if both xApps simultaneously issued control commands to the same \gls{ran} E2 nodes. For instance, if both versions simultaneously commanded different PRB allocation ratios for the same UE, the base station would experience rapid parameter changes every few seconds, potentially leading to scheduler instability or degraded network performance. This control conflict prevention is specific to RAN environments and differs from typical cloud-native A/B testing where services operate independently.

\begin{figure}[!ht]
\centering
\includegraphics[width=\columnwidth]{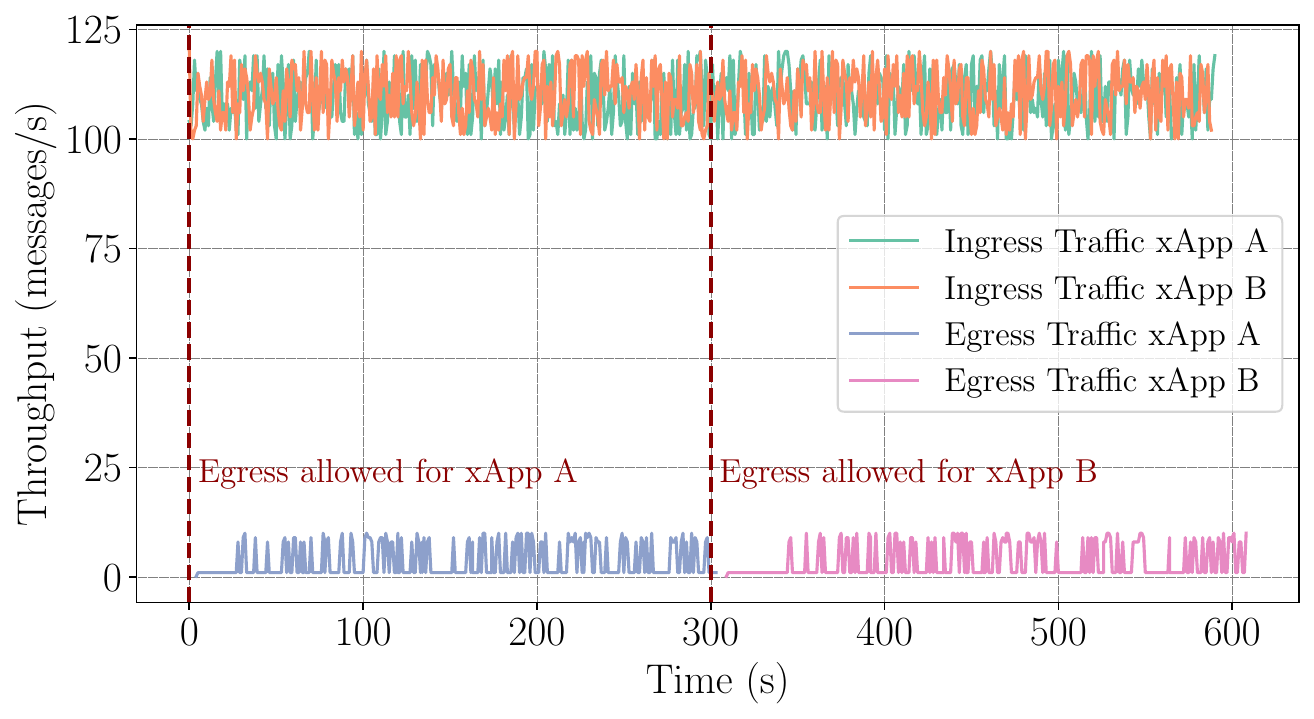}
\caption{A/B test for xApp version A and xApp version B.}
\label{fig:ab-testing}
\end{figure}

This experiment was structured over a 10-minute period. During the first 5 minutes, both xApps receive ingress traffic, but only xApp A is permitted to send egress control messages, thus operating as the active controller, while xApp B runs in observation-only mode. After 5 minutes, Istio circuit breaking transfers control authority from xApp A to B by blocking the former xApp's egress traffic and enabling that of the latter for the remaining 5 minutes.
Figure~\ref{fig:ab-testing} shows the traffic patterns during the A/B test.  We notice that ingress traffic remains balanced at approximately $100$\:msg/s for both xApp versions throughout the whole experiment, ensuring symmetric observational data. The egress traffic patterns demonstrate the circuit-breaking mechanism in action. During the first 5 minutes, the egress of xApp~A mirrors its ingress, while that of xApp~B remains at $0$. At the 5-minute transition point, a reversal occurs: the egress of xApp~A drops to $0$, while that of xApp~B starts to flow. The precision of this switchover, with no observable overlap and just a slight gap of $6$\:s in control authority due to the Ansible playbook action delay, validates the effectiveness of the circuit-breaking implementation used by \oursol. This mechanism is key in enabling safe production testing without risking control conflicts or network instability.

%%%%%%%%%%%%%%%%%%%%%%%%%%%%%%%%%%%%%%%%%%%%%%%%%%%%%%%%%%%%%%%%%%%%%%%%%%%%%%%%%%%%%%%%%%%%%%%%%%%%%%%%%%%%%%%%%%
%%%%%%%%%%%%%%%%%%%%%%%%%%%%%%%%%%%%%%%%%%%%%%%%%%%%%%%%%%%%%%%%%%%%%%%%%%%%%%%%%%%%%%%%%%%%%%%%%%%%%%%%%%%%%%%%%%
%CONCLUSIONS
%%%%%%%%%%%%%%%%%%%%%%%%%%%%%%%%%%%%%%%%%%%%%%%%%%%%%%%%%%%%%%%%%%%%%%%%%%%%%%%%%%%%%%%%%%%%%%%%%%%%%%%%%%%%%%%%%%
%%%%%%%%%%%%%%%%%%%%%%%%%%%%%%%%%%%%%%%%%%%%%%%%%%%%%%%%%%%%%%%%%%%%%%%%%%%%%%%%%%%%%%%%%%%%%%%%%%%%%%%%%%%%%%%%%%

\section{Conclusions and Future Work}\label{sec:conclusions}
The evolution of Open \gls{ran} architectures has introduced unprecedented opportunities for innovation through open interfaces and disaggregated components, yet the practical deployment and testing of xApps remains hindered by significant operational challenges. Indeed, traditional \gls{ran} environments lack the DevOps practices and tooling necessary to support continuous integration, automated testing, and reliable lifecycle management of xApps across diverse deployment contexts.

In this paper, we introduced \oursol, our original DevOps platform built on \gls{cicd} and service mesh technologies that addresses these critical gaps. \oursol enables comprehensive lifecycle management of xApps within the \nearrtric through automated \gls{cicd} pipelines, providing fine-grained control over traffic management, observability, and security. Our experimental validation across simulated, emulated, and real testbed environments demonstrates the capabilities of \oursol to streamline xApp orchestration while enabling advanced deployment strategies such as canary releases and A/B testing. We believe this work significantly improves xApp workload management in \gls{ran} environments, enabling extensive testing while avoiding unexpected control behavior on production E2 nodes.

In line with latest community initiatives such as those of the AI-\gls{ran} Alliance, future evolutions of \oursol will include adding support for MLOps xApp workflows, simplifying service mesh configuration through emerging technologies such as \gls{mcp}, investigating in-kernel acceleration techniques using P4 and eBPF, as well as architectural extensions to support rApps and dApps.

%%%%%%%%%%%%%%%%%%%%%%%%%%%%%%%%%%%%%%%%%%%%%%%%%%%%%%%%%%%%%%%%%%%%%%%%%%%%%%%%%%%%%%%%%%%%%%%%%%%%%%%%%%%%%%%%%%
%%%%%%%%%%%%%%%%%%%%%%%%%%%%%%%%%%%%%%%%%%%%%%%%%%%%%%%%%%%%%%%%%%%%%%%%%%%%%%%%%%%%%%%%%%%%%%%%%%%%%%%%%%%%%%%%%%
%ACK
%%%%%%%%%%%%%%%%%%%%%%%%%%%%%%%%%%%%%%%%%%%%%%%%%%%%%%%%%%%%%%%%%%%%%%%%%%%%%%%%%%%%%%%%%%%%%%%%%%%%%%%%%%%%%%%%%%
%%%%%%%%%%%%%%%%%%%%%%%%%%%%%%%%%%%%%%%%%%%%%%%%%%%%%%%%%%%%%%%%%%%%%%%%%%%%%%%%%%%%%%%%%%%%%%%%%%%%%%%%%%%%%%%%%%

\section*{Acknowledgments}

This work has been partially supported by the National Telecommunications and Information Administration (NTIA)'s Public Wireless Supply Chain Innovation Fund (PWSCIF) under Award No. 25-60-IF054, and by the European Union under the Italian National Recovery and Resilience Plan (NRRP) of NextGenerationEU, partnership on “Telecommunications of the Future” (PE00000001 - program “RESTART”) CUP: J33C22002880001.

\bibliographystyle{IEEEtran}
\bibliography{bib}

\begin{thebibliography}{10}
\providecommand{\url}[1]{#1}
\csname url@samestyle\endcsname
\providecommand{\newblock}{\relax}
\providecommand{\bibinfo}[2]{#2}
\providecommand{\BIBentrySTDinterwordspacing}{\spaceskip=0pt\relax}
\providecommand{\BIBentryALTinterwordstretchfactor}{4}
\providecommand{\BIBentryALTinterwordspacing}{\spaceskip=\fontdimen2\font plus
\BIBentryALTinterwordstretchfactor\fontdimen3\font minus
  \fontdimen4\font\relax}
\providecommand{\BIBforeignlanguage}[2]{{%
\expandafter\ifx\csname l@#1\endcsname\relax
\typeout{** WARNING: IEEEtran.bst: No hyphenation pattern has been}%
\typeout{** loaded for the language `#1'. Using the pattern for}%
\typeout{** the default language instead.}%
\else
\language=\csname l@#1\endcsname
\fi
#2}}
\providecommand{\BIBdecl}{\relax}
\BIBdecl

\bibitem{understandingoran2023polese}
M.~Polese, L.~Bonati, S.~D'Oro, S.~Basagni, and T.~Melodia, ``Understanding
  o-ran: Architecture, interfaces, algorithms, security, and research
  challenges,'' \emph{IEEE Communications Surveys \& Tutorials}, vol.~25,
  no.~2, pp. 1376--1411, 2023.

\bibitem{wg1}
\BIBentryALTinterwordspacing
{O-RAN Alliance, Alfter Germany}, ``{O-RAN Architecture Description,
  O-RAN.WG1.TS.OAD-R004-v13.00},'' {O-RAN Alliance}, Technical Specification
  (TS) 13.00, 2024. [Online]. Available:
  \url{https://specifications.o-ran.org/download?id=789}
\BIBentrySTDinterwordspacing

\bibitem{tutorial5g2025}
J.~L. Herrera, S.~Montebugnoli, D.~Scotece, L.~Foschini, and P.~Bellavista, ``A
  tutorial on o-ran deployment solutions for 5g: From simulation to emulated
  and real testbeds,'' \emph{IEEE Communications Surveys \& Tutorials}, pp.
  1--1, 2025.

\bibitem{maxenti2025autoranautomatedzerotouchopen}
\BIBentryALTinterwordspacing
S.~Maxenti, R.~Shirkhani, M.~Elkael, L.~Bonati, S.~D'Oro, T.~Melodia, and
  M.~Polese, ``Autoran: Automated and zero-touch open ran systems,'' 2025.
  [Online]. Available: \url{https://arxiv.org/abs/2504.11233}
\BIBentrySTDinterwordspacing

\bibitem{fromzerotohero2025}
J.~F. Santos, A.~Huff, D.~Campos, K.~V. Cardoso, C.~B. Both, and L.~A. DaSilva,
  ``Managing o-ran networks: xapp development from zero to hero,'' \emph{IEEE
  Communications Surveys \& Tutorials}, pp. 1--1, 2025.

\bibitem{TIFG}
\BIBentryALTinterwordspacing
{O-RAN Alliance, Alfter Germany}, ``{O-RAN End-to-end Test Specification 7.0,
  O-RAN.TIFG.TS.E2E-Test.0-R004-v07.00},'' {O-RAN Alliance}, Technical
  Specification (TS) 2.00, 2025. [Online]. Available:
  \url{https://specifications.o-ran.org/download?id=922}
\BIBentrySTDinterwordspacing

\bibitem{kim2016devops}
G.~Kim, J.~Humble, P.~Debois, and J.~Willis, \emph{The DevOps Handbook: How to
  Create World-Class Agility, Reliability, and Security in Technology
  Organizations}.\hskip 1em plus 0.5em minus 0.4em\relax IT Revolution Press,
  2016.

\bibitem{intelligentoranbeyond5g2024}
S.~Marinova and A.~Leon-Garcia, ``Intelligent o-ran beyond 5g: Architecture,
  use cases, challenges, and opportunities,'' \emph{IEEE Access}, vol.~12, pp.
  27\,088--27\,114, 2024.

\bibitem{wg6}
\BIBentryALTinterwordspacing
{O-RAN Alliance, Alfter Germany}, ``{Cloud Platform Reference Designs,
  O-RAN.WG6.CADS-v08.01},'' {O-RAN Alliance}, Technical Report (TR) 8.01, 2024.
  [Online]. Available: \url{https://specifications.o-ran.org/download?id=817}
\BIBentrySTDinterwordspacing

\bibitem{colosseum2021}
L.~Bonati, P.~Johari, M.~Polese, S.~D'Oro, S.~Mohanti, M.~Tehrani-Moayyed,
  D.~Villa, S.~Shrivastava, C.~Tassie, K.~Yoder, A.~Bagga, P.~Patel, V.~Petkov,
  M.~Seltser, F.~Restuccia, A.~Gosain, K.~R. Chowdhury, S.~Basagni, and
  T.~Melodia, ``Colosseum: Large-scale wireless experimentation through
  hardware-in-the-loop network emulation,'' in \emph{2021 IEEE International
  Symposium on Dynamic Spectrum Access Networks (DySPAN)}, 2021, pp. 105--113.

\bibitem{xdevsmferaudo2024}
\BIBentryALTinterwordspacing
A.~Feraudo, S.~Maxenti, A.~Lacava, P.~Bellavista, M.~Polese, and T.~Melodia,
  ``xdevsm: Streamlining xapp development with a flexible framework for o-ran
  e2 service models,'' in \emph{Proceedings of the 30th Annual International
  Conference on Mobile Computing and Networking}, ser. ACM MobiCom '24.\hskip
  1em plus 0.5em minus 0.4em\relax New York, NY, USA: Association for Computing
  Machinery, 2024, p. 1954–1961. [Online]. Available:
  \url{https://doi.org/10.1145/3636534.3697325}
\BIBentrySTDinterwordspacing

\bibitem{cncfmesh}
{Cloud Native Computing Foundation (CNCF)}, ``{CNCF Documentation on Service
  Mesh},''
  \url{https://landscape.cncf.io/guide#orchestration-management--service-mesh},
  2025.

\bibitem{lookatservicemesh}
A.~Koschel, M.~Bertram, R.~Bischof, K.~Schulze, M.~Schaaf, and I.~Astrova, ``{A
  Look at Service Meshes},'' in \emph{2021 12th International Conference on
  Information, Intelligence, Systems \& Applications (IISA)}, 2021, pp. 1--8.

\bibitem{3gpp2023ts23501}
\BIBentryALTinterwordspacing
{3GPP}, ``System architecture for the 5g system (5gs); stage 2,'' 3rd
  Generation Partnership Project, Technical Specification TS 23.501, June 2023,
  release 18. [Online]. Available:
  \url{https://www.3gpp.org/ftp/Specs/archive/23_series/23.501/}
\BIBentrySTDinterwordspacing

\bibitem{challengesopportunitiesservicemesh}
W.~Li, Y.~Lemieux, J.~Gao, Z.~Zhao, and Y.~Han, ``{Service Mesh: Challenges,
  State of the Art, and Future Research Opportunities},'' in \emph{2019 IEEE
  International Conference on Service-Oriented System Engineering (SOSE)},
  2019, pp. 122--1225.

\bibitem{soldanisauron2023}
D.~Soldani, P.~Nahi, H.~Bour, S.~Jafarizadeh, M.~F. Soliman, L.~Di~Giovanna,
  F.~Monaco, G.~Ognibene, and F.~Risso, ``{eBPF: A New Approach to Cloud-Native
  Observability, Networking and Security for Current (5G) and Future Mobile
  Networks (6G and Beyond)},'' \emph{IEEE Access}, vol.~11, pp.
  57\,174--57\,202, 2023.

\bibitem{10719389}
V.~Chawla, ``{Exploring How Service Meshes Can Help To Facilitate The
  Development Of Microservices},'' in \emph{2024 IEEE 15th International
  Conference on Software Engineering and Service Science (ICSESS)}, 2024, pp.
  35--43.

\bibitem{10.1007/978-3-030-33702-5_12}
H.~Johng, A.~K. Kalia, J.~Xiao, M.~Vukovi{\'{c}}, and L.~Chung, ``{Harmonia: A
  Continuous Service Monitoring Framework Using DevOps and Service Mesh in a
  Complementary Manner},'' in \emph{Service-Oriented Computing}, S.~Yangui,
  I.~Bouassida~Rodriguez, K.~Drira, and Z.~Tari, Eds.\hskip 1em plus 0.5em
  minus 0.4em\relax Cham: Springer International Publishing, 2019, pp.
  151--168.

\bibitem{10.1145/3194760.3194763}
\BIBentryALTinterwordspacing
T.~F. D\"{u}llmann, C.~Paule, and A.~van Hoorn, ``{Exploiting devops practices
  for dependable and secure continuous delivery pipelines},'' in
  \emph{Proceedings of the 4th International Workshop on Rapid Continuous
  Software Engineering}, ser. RCoSE '18.\hskip 1em plus 0.5em minus 0.4em\relax
  New York, NY, USA: Association for Computing Machinery, 2018, p. 27–30.
  [Online]. Available: \url{https://doi.org/10.1145/3194760.3194763}
\BIBentrySTDinterwordspacing

\bibitem{servicemeshreadiness5g2023}
S.~Aldas and A.~Babakian, ``Cloud-native service mesh readiness for 5g and
  beyond,'' \emph{IEEE Access}, vol.~11, pp. 132\,286--132\,295, 2023.

\bibitem{duongservicemeshcore5g2023}
V.-B. Duong and Y.~Kim, ``A design of service mesh based 5g core network using
  cilium,'' in \emph{2023 International Conference on Information Networking
  (ICOIN)}, 2023, pp. 25--28.

\bibitem{ambient}
\BIBentryALTinterwordspacing
{Istio Authors}, ``{Istio Ambient Mesh Documentation}.'' [Online]. Available:
  \url{https://istio.io/latest/blog/2022/introducing-ambient-mesh/}
\BIBentrySTDinterwordspacing

\bibitem{cilium}
\BIBentryALTinterwordspacing
{Cilium Authors}, ``{Cilium},'' 2025. [Online]. Available:
  \url{https://docs.cilium.io/}
\BIBentrySTDinterwordspacing

\bibitem{istio}
\BIBentryALTinterwordspacing
{Istio Authors}, ``Istio documentation,'' 2025. [Online]. Available:
  \url{https://istio.io/latest/about/service-mesh/}
\BIBentrySTDinterwordspacing

\bibitem{zhu2022dissecting}
X.~Zhu, G.~She, B.~Xue, Y.~Zhang, Y.~Zhang, X.~K. Zou, X.~Duan, P.~He,
  A.~Krishnamurthy, M.~Lentz, D.~Zhuo, and R.~Mahajan, ``{Dissecting Service
  Mesh Overheads},'' 2022.

\bibitem{servicemesh_ebpf}
M.~R.~S. Sedghpour and P.~Townend, ``{Service Mesh and eBPF-Powered
  Microservices: A Survey and Future Directions},'' in \emph{2022 IEEE
  International Conference on Service-Oriented System Engineering (SOSE)},
  2022, pp. 176--184.

\bibitem{servicemeshmeta}
\BIBentryALTinterwordspacing
H.~Saokar, S.~Demetriou, N.~Magerko, M.~Kontorovich, J.~Kirstein, M.~Leibold,
  D.~Skarlatos, H.~Khandelwal, and C.~Tang, ``{ServiceRouter}: Hyperscale and
  minimal cost service mesh at meta,'' in \emph{17th USENIX Symposium on
  Operating Systems Design and Implementation (OSDI 23)}.\hskip 1em plus 0.5em
  minus 0.4em\relax Boston, MA: USENIX Association, July 2023, pp. 969--985.
  [Online]. Available:
  \url{https://www.usenix.org/conference/osdi23/presentation/saokar}
\BIBentrySTDinterwordspacing

\bibitem{5gct2025}
L.~Bonati, M.~Polese, S.~D'Oro, P.~B. del Prever, and T.~Melodia, ``5g-ct:
  Automated deployment and over-the-air testing of end-to-end open radio access
  networks,'' \emph{IEEE Communications Magazine}, vol.~63, no.~1, pp.
  155--160, 2025.

\bibitem{orchestran2022}
{D'Oro, Salvatore and Bonati, Leonardo and Polese, Michele and Melodia,
  Tommaso}, ``{OrchestRAN: Network Automation through Orchestrated Intelligence
  in the Open RAN},'' in \emph{{IEEE INFOCOM 2022 - IEEE Conference on Computer
  Communications}}, 2022, pp. 270--279.

\bibitem{9647628}
K.~Trantzas, C.~Tranoris, S.~Denazis, R.~Direito, D.~Gomes, J.~Gallego-Madrid,
  A.~Hermosilla, and A.~Skarmeta, ``An automated ci/cd process for testing and
  deployment of network applications over 5g infrastructure,'' in \emph{2021
  IEEE International Mediterranean Conference on Communications and Networking
  (MeditCom)}, 2021, pp. 156--161.

\bibitem{9376232}
B.~Balasubramanian, E.~S. Daniels, M.~Hiltunen, R.~Jana, K.~Joshi, R.~Sivaraj,
  T.~X. Tran, and C.~Wang, ``Ric: A ran intelligent controller platform for
  ai-enabled cellular networks,'' \emph{IEEE Internet Computing}, vol.~25,
  no.~2, pp. 7--17, 2021.

\bibitem{10329915}
M.~Hoffmann, S.~Janji, A.~Samorzewski, Å.~Kułacz, C.~Adamczyk, M.~Dryjański,
  P.~Kryszkiewicz, A.~Kliks, and H.~Bogucka, ``Open ran xapps design and
  evaluation: Lessons learnt and identified challenges,'' \emph{IEEE Journal on
  Selected Areas in Communications}, vol.~42, no.~2, pp. 473--486, 2024.

\bibitem{9681936}
H.~Lee, Y.~Jang, J.~Song, and H.~Yeon, ``O-ran ai/ml workflow implementation of
  personalized network optimization via reinforcement learning,'' in \emph{2021
  IEEE Globecom Workshops (GC Wkshps)}, 2021, pp. 1--6.

\bibitem{10858200}
A.~Da~Silva, M.~R. Chowdhury, A.~Sathish, A.~Tripathi, S.~F. Midkiff, and L.~A.
  Da~Silva, ``Cci xg testbed: An o-ran based platform for future wireless
  network experimentation,'' \emph{IEEE Communications Magazine}, vol.~63,
  no.~2, pp. 62--68, 2025.

\bibitem{10.1145/3570361.3615745}
\BIBentryALTinterwordspacing
P.~Bahl, M.~Balkwill, X.~Foukas, A.~Kalia, D.~Kim, M.~Kotaru, Z.~Lai,
  S.~Mehrotra, B.~Radunovic, S.~Saroiu, C.~Settle, A.~Verma, A.~Wolman, F.~Y.
  Yan, and Y.~Zhang, ``Accelerating open ran research through an
  enterprise-scale 5g testbed,'' in \emph{Proceedings of the 29th Annual
  International Conference on Mobile Computing and Networking}, ser. ACM
  MobiCom '23.\hskip 1em plus 0.5em minus 0.4em\relax New York, NY, USA:
  Association for Computing Machinery, 2023. [Online]. Available:
  \url{https://doi.org/10.1145/3570361.3615745}
\BIBentrySTDinterwordspacing

\bibitem{10901151}
S.~Montebugnoli, A.~Sabbioni, and L.~Foschini, ``Evaluating mesh communications
  in disaggregated near-rt ric for 5g open ran: a functional and performance
  analysis,'' in \emph{GLOBECOM 2024 - 2024 IEEE Global Communications
  Conference}, 2024, pp. 571--576.

\bibitem{10327727}
S.~Aldas and A.~Babakian, ``Cloud-native service mesh readiness for 5g and
  beyond,'' \emph{IEEE Access}, vol.~11, pp. 132\,286--132\,295, 2023.

\bibitem{fi17080372}
\BIBentryALTinterwordspacing
M.~Hashem~Eiza, B.~Akwirry, A.~Raschella, M.~Mackay, and M.~K. Maheshwari, ``A
  hybrid zero trust deployment model for securing o-ran architecture in 6g
  networks,'' \emph{Future Internet}, vol.~17, no.~8, 2025. [Online].
  Available: \url{https://www.mdpi.com/1999-5903/17/8/372}
\BIBentrySTDinterwordspacing

\bibitem{k3s}
\BIBentryALTinterwordspacing
{Rancher Labs}, ``K3s: Lightweight kubernetes,'' 2025. [Online]. Available:
  \url{https://k3s.io/}
\BIBentrySTDinterwordspacing

\bibitem{oransc}
\BIBentryALTinterwordspacing
{O-RAN Software Community}, ``{O-RAN Software Community},'' 2025. [Online].
  Available: \url{https://github.com/o-ran-sc}
\BIBentrySTDinterwordspacing

\bibitem{cncf}
\BIBentryALTinterwordspacing
{Cloud Native Computing Foundation and Linux Foundation}, ``{The Cloud Native
  Computing Foundation Landscape},'' 2025, accessed: 2025-05-29. [Online].
  Available:
  \url{https://landscape.cncf.io/?view-mode=card&classify=category&sort-by=name&sort-direction=asc#orchestration-management--service-mesh}
\BIBentrySTDinterwordspacing

\bibitem{RODRIGUES2025112500}
\BIBentryALTinterwordspacing
H.~Rodrigues, A.~{Rito Silva}, and A.~Avritzer, ``Assessment of performance and
  its scalability in microservice architectures: Systematic literature
  review,'' \emph{Journal of Systems and Software}, vol. 230, p. 112500, 2025.
  [Online]. Available:
  \url{https://www.sciencedirect.com/science/article/pii/S0164121225001682}
\BIBentrySTDinterwordspacing

\end{thebibliography}

\begin{IEEEbiographynophoto}
{Sofia Montebugnoli}
(Student Member, IEEE) is a Ph.D. student in Computer Science and Engineering from the University of Bologna. She received a Master's Degree in Computer Engineering from the University of Bologna in 2022.  Her main research interests include O-RAN, cloud continuum, and middleware.

\vspace{5pt}
\noindent
\textbf{Leonardo Bonati} (Member, IEEE) is an Associate Research Scientist at the Institute for the Wireless Internet of Things at Northeastern University. He received a Ph.D. degree in Computer Engineering from Northeastern University in 2022. His research focuses on softwarized Open RAN, network automation, orchestration, and virtualization for next-generation cellular networks. He received the 2024 Mario Gerla Award for Research in Computer Science. He served as TPC co-chair for IEEE DTwin 2025, co-chair for the Testbeds and Experimentation track at IEEE CCNC 2025 and 2026, and guest editor for a special issue of Elsevier Computer Networks on experimental wireless platforms.

\vspace{5pt}
\noindent
\textbf{Andrea Sabbioni} (Member, IEEE) received his PhD in Computer Science and Engineering in 2022, and he is currently a Junior Researcher at the University of Bologna (RTD-A), Italy. His research interests include management and orchestration in the Cloud Continuum and Serverless Computing.

\vspace{5pt}
\noindent
\textbf{Luca Foschini} (Senior Member, IEEE) received a Ph.D. degree in computer science engineering from the University of Bologna, Italy, where he is now full professor of distributed systems. His interests span from integrated management of distributed systems and services to mobile crowd-sourcing/-sensing, from infrastructures for Industry 4.0 to fog/edge cloud systems.

\vfill\eject

\vspace{5pt}
\noindent
\textbf{Paolo Bellavista} (Senior Member, IEEE) is professor of distributed and mobile systems at the University of Bologna, Italy. His research activities span from pervasive wireless computing to middleware for the cloud continuum, from ultra-low-latency support in Industry 4.0 applications to digital twins of industrial processes and smart cities. He serves in several Editorial Boards of IEEE/ACM/Elsevier international journals, including serving as Associate EiC of IEEE COMST.

\vspace{5pt}
\noindent
\textbf{Salvatore D'Oro} (Senior Member, IEEE) is the CTO and co-founder of zTouch Networks, a company focused on the development of zero-touch automation solutions for O-RAN systems. He is also a Research Associate Professor at Northeastern University. He received his Ph.D. degree from the University of Catania and is an area editor of Elsevier Computer Communications journal. He serves on the TPC of IEEE INFOCOM, IEEE CCNC \& ICC and IFIP Networking. He is one of the contributors to OpenRAN Gym, the first open-source research platform for AI/ML applications in the Open RAN. His research interests include optimization, AI \& network slicing for NextG Open RANs.

\vspace{5pt}
\noindent
\textbf{Michele Polese} (Senior Member, IEEE) is a Research Assistant Professor at the Institute for the Wireless Internet of Things, Northeastern University. He earned his Ph.D. from the University of Padova in 2020 and held research roles at NYU, AT\&T Labs, and Northeastern. His work focuses on 5G/6G networks, mmWave/THz communication, Open RAN, and spectrum sharing. He has contributed to O-RAN specs, FCC/NTIA consultations, serves on the AMS Committee on RF Allocations, and since July 2025 he is chairing the AI-RAN Alliance AI-and-RAN working group. He is PI/co-PI on projects funded by NTIA, NSF, and others, and received the 2022 Mario Gerla Award. He also serves as editor, organizer, and TPC chair in major IEEE venues.

\vspace{5pt}
\noindent
\textbf{Tommaso Melodia} (Fellow, IEEE) is the William Lincoln Smith Chair Professor at Northeastern University and Founding Director of the Institute for the Wireless Internet of Things. He also serves as Director of Research for the \$100M PAWR Project Office. He received his Ph.D. from Georgia Tech in 2007 and is a recipient of the NSF CAREER award. His research focuses on IoT and wireless networked systems and is funded by NSF, DARPA, ONR, AFRL, and ARL. He is an IEEE Fellow, ACM Distinguished Member, and has held editorial and leadership roles in major conferences, including IEEE INFOCOM and ACM Nanocom.
\end{IEEEbiographynophoto}

\end{document}